\documentstyle[12pt]{article}

\renewcommand{\theequation}{\thesection.\arabic{equation}}

\def\kph{{ KP hierarchy}}

\def\tauf{ the $\tau$--function}

\def\kdvh{{KdV hierarchy}}
\def\kdvhs{{KdV hierarchies}}

\def\gkdvhs{higher KdV hierarchies}
\def\bhs{bi--Hamiltonian structure}
\def\shs{second Hamiltonian structure}

\def\a{\begin{eqnarray}}
\def\b{\end{eqnarray}}
\def\0{\nonumber}
\def\ba{\begin{array}}
\def\ea{\end{array}}
\def\noal{\noalign{\vskip10pt}}

\def\ha{{1\over2}}
\def\al{{\alpha}}
\def\lm{{\lambda}}
\def\dxy{\delta(x-y)}
\def\ddxy{\delta'(x-y)}

\newlength{\extraspace}
\newlength{\extraspaces}
\newcounter{dummy}

\newcommand{\ai}{
\addtocounter{equation}{1}
\setcounter{dummy}{\value{equation}}
\setcounter{equation}{0}
\renewcommand{\theequation}{\thesection.\arabic{dummy}\alph{equation}}
\begin{eqnarray}
\addtolength{\abovedisplayskip}{\extraspaces}
\addtolength{\belowdisplayskip}{\extraspaces}
\addtolength{\abovedisplayshortskip}{\extraspace}
\addtolength{\belowdisplayshortskip}{\extraspace}}
\newcommand{\bj}{
\end{eqnarray}
\setcounter{equation}{\value{dummy}}
\renewcommand{\theequation}{\thesection.\arabic{equation}}}

\def\d{{\partial}}

\def\lm{{\lambda}}

\newcommand{\ddx}{{\partial \over \partial x}}

\newcommand{\ddt}[1]{{\partial \over \partial t_{#1}}}

\newcommand{\inv}{^{-1}}

\newcommand{\th}{^{\mit th}}
\newcommand{\res}{{\rm res}\,}

\newcommand{\bac}{\begin{array}{c}}
\newcommand{\bacc}{\begin{array}{cc}}
\newcommand{\baccc}{\begin{array}{ccc}}
\newcommand{\barcl}{\begin{array}{rcl}}
\newcommand{\bacccc}{\begin{array}{cccc}}
\newcommand{\baccccc}{\begin{array}{ccccc}}
\newcommand{\baccccccc}{\begin{array}{ccccccc}}
\newcommand{\barclcrcl}{\begin{array}{rclcrcl}}
\newcommand{\bacl}{\begin{array}{cl}}
\newcommand{\bacll}{\begin{array}{cll}}
\newcommand{\eac}{\end{array}}

\newcounter{tabnum}
\begin{document}

\begin{flushright}
SISSA-171/93/EP\\
BONN--HE--46/93\\
AS--ITP--49/93\\
hep-th/93mmxxx
\end{flushright}
\vskip0.5cm

\centerline{\LARGE\bf The $(N, M)$--th \kdvh}
\vskip.5cm
\centerline{\LARGE\bf and the associated $W$ algebra}
\vskip1.0cm
\centerline{\large  L. Bonora}
\vskip.4cm
\centerline{International School for Advanced Studies (SISSA/ISAS)}
\centerline{Via Beirut 2, 34014 Trieste, Italy}
\centerline{INFN, Sezione di Trieste}
\vskip.7cm
\centerline{\large  C. S. Xiong}
\vskip.4cm
\centerline{Physikalisches Institut, der Universit\"at Bonn}
\centerline{Nussallee 12, 53115 Bonn, Germany}
\vskip.4cm
\centerline{Institute of Theoretical Physics, Academia Sinica}
\centerline{P. O. Box 2735, Beijing 100080, China}
\vskip2cm
\abstract{We discuss a differential integrable hierarchy,
which we call the $(N, M)$--th \kdvh, whose Lax operator
is obtained by properly adding $M$ pseudo--differential terms to the
Lax operator of the $N$--th KdV hierarchy. This new hierarchy contains both the
higher \kdvh\, and multi--field representation of \kph\, as sub--systems and
naturally appears in multi--matrix models. The $N+2M-1$ coordinates
or fields of this hierarchy satisfy two algebras of compatible Poisson
brackets which are {\it local} and {\it polynomial}. Each Poisson
structure generate an extended $W_{1+\infty}$ and $W_\infty$ algebra,
respectively. We call $W(N, M)$ the generating algebra of the extended
$W_\infty$ algebra. This algebra, which corresponds with the second
Poisson structure, shares many features of the usual $W_N$
algebra. We show that there exist $M$ distinct reductions
of the $(N, M)$--th \kdvh, which are obtained
by imposing suitable second class constraints. The most drastic reduction
corresponds to the $(N+M)$--th \kdvh. Correspondingly the $W(N, M)$
algebra is reduced to the $W_{N+M}$ algebra. We study
in detail the dispersionless limit of this
hierarchy and the relevant reductions.}
\vfill
\eject

\section{Introduction}

\setcounter{equation}{0}
\setcounter{subsection}{0}
\setcounter{footnote}{0}

Integrable hierarchies are a topic of increasing importance
in theoretical physics. It has been realized
recently that they
play an essential role in the study of 2D quantum gravity and
topological field theories, as well as in matrix models. In particular
matrix models exhibit an extremely rich integrable structure.
The origin of this interest in integrable hierarchies
goes back to 1989 when three different groups, through the so--called
double scaling limit technique, obtained remarkable non--perturbative
results in 2D quantum gravity. In particular they found
that the partition function of one--hermitean matrix model with
even potential satisfies
the KdV hierarchy equations \cite{gross}. Later on it has been shown that
topological field theories coupled to topological gravity and
the Kontsevich model also possess this integrable
structure\cite{w}\cite{k}.
After these successes several authors have conjectured that
multi--matrix models should be governed by higher \kdvhs\cite{d}.
Now, it is rather straightforward to extract discrete hierarchies from matrix
models. The difficulties start when we try to pass to differential hierarchies.
In particular the double scaling limit technique does not prove as
powerful and manageable in multi--matrix models as in one--matrix
models.

In \cite{bx1} we proposed an alternative approach to investigate one--matrix
models, by which we could extract a differential integrable hierarchy from the
discrete
one {\it without reference to any continuum limit}.
Our basic observation is the following one:
in one--hermitean matrix model there naturally
exists a discrete (or lattice) integrable hierarchy -- the Toda lattice
hierarchy; if we treat the first flow parameter as {\it space coordinate}, this
discrete integrable hierarchy can be re--expressed as a {\it continuum}\,
(differential) integrable hierarchy, which admits the following Lax pair
representation
\ai
&&L_2=\d+R\frac{1}{\d-S},\label{l2}\\
&&\ddt r L_2=[(L_2^r)_+, L_2]\label{kp2}
\bj
where $\d=\frac{\partial}{\partial x}=\frac{\partial}{\partial t_1}$,
$t_1$ is the space coordinate, while $t_r(r\geq2)$ are real `time'
(coupling) parameters.
The subscript ``$+$" means that we keep only the terms containing
non--negative powers of $\d$.
$R$ and $S$ are the independent fields of the system.
This differential integrable hierarchy is referred to as two--boson
representation of the \kph\footnote{This integrable structure also shows up
in WZW model and Conformal Affine Toda field theories (CAT models)\cite{yw}.
For a more mathematical approach see \cite{manas}  and references therein}.
This hierarchy can be reduced
to the \kdvh\ by imposing the second class constraint $S=0$.

In \cite{bx2} we showed that this approach is also applicable to multi--matrix
models (with bilinear couplings among different matrices) and permits a
systematic analysis of these models. Such multi--matrix models are
characterized by generalized Toda lattice integrable hierarchies which,
via the same procedure applied to one--matrix models,
can be rewritten as the following differential integrable hierarchies
\ai
&&L_{2M}=\d +\sum_{l=1}^{M} a_l \frac{1}{\d -S_l}
  \frac{1}{\d -S_{l-1}}\cdots \frac{1}{\d -S_1}\label{l2m}\\
&&\ddt r L_{2M}= [ (L_{2M}^r)_+, L_{2M}] \label{kp2m}
\bj
where $a_1,..., a_{M}, S_1,..., S_{M}$  are independent coordinates
({\it fundamental fields} or {\it dynamical variables}) of the system.
These integrable hierarchies are called {\it 2M--field representations of
the \kph}. It is just these differential integrable hierarchies (together
with the string equations) that contain the information of
multi--matrix models. In \cite{bx3} we showed that they are related to the
\gkdvhs \, via Hamiltonian reduction.

In particular in \cite{bx3} we examined in detail
the $4$--field representation of the \kph\ and we obtained two distinct
integrable hierarchies by suppressing successively the fields $S_1, S_2$.
The corresponding Lax operators are
\a
L_{21}=\d^2+a_1+a_2\frac{1}{\d-S_2},\qquad
L_{3}=\d^3+a_1\d+a_2.\0
\b

In the present paper we carry our analysis further along the same line.
Precisely we will show that:

\noindent
$(i)$. There exists a general integrable differential hierarchy
\a
\ddt r L=[\Bigl(L^{r\over{N}}\Bigl)_+, L] \label{gkdvh}
\b
with the following Lax operator
\a
L=\d^{N}+\sum_{l=1}^{N-1} a_l\d^{N-l-1}
 +\sum_{l=1}^M a_{N+l-1}{1\over{\d-S_l}}{1\over{\d-S_{l-1}}}
  \ldots {1\over{\d-S_1}},\quad N\ge1,\quad M\ge0; \label{pdo}
\b
which involves $(N+2M-1)$ independent fields. We call them the
coordinates of the hierarchy.
Since the Lax operator (\ref{pdo}) is obtained by adding
$M$ pseudo--differential terms to the Lax operator of the usual $N$--th \kdvh,
and in the lack of a preexisting terminology, we refer to this integrable
hierarchy as the $(N, M)$--th KdV hierarchy (or the $M$--extended
$N$--th KdV hierarchy). We refer at times to the $N+2M-1$ fields together
with all the properties implied by integrability as `{\it model}'. So,
for example, we will be talking about the $(N,M)$ model in connection with
the hierarchy just introduced.

\noindent
$(ii)$. Secondly we will show that the coordinates of the $(N, M)$--th \kdvh\,
form two closed algebras with respect to the two Poisson brackets which
constitute the bi--Hamiltonian structure. They generate  the extended
$W_{1+\infty}$ and $W_\infty$ algebras, respectively: in other words, there
exist combinations of the
fields and their derivatives which satisfy the extended $W_{1+\infty}$ and
$W_\infty$ algebras.

\noindent
$(iii)$. Finally we will prove that there exist various
Hamiltonian reductions.  Precisely we show that it is possible
to suppress the $S$ fields
one by one and still obtain integrable hierarchies; at the end
of this cascade reduction one gets a KdV hierarchy of order $N+M$.

The motivation for this work stems from the fact that the pseudodifferential
operator (\ref{pdo}) is, after reduction to the standard form,
the most general operator which appears in multi--matrix models.
In a companion paper \cite{bxII} we use the hierarchies of this paper
to calculate the correlation functions of the corresponding matrix models.

We remark that there is not only one possible viewpoint to study the
system specified by (\ref{gkdvh}) and (\ref{pdo}). In fact, a posteriori,
one will remark that the Lax operator (\ref{pdo}) can be envisaged as a
reduction of Lax operator of the type (\ref{l2m}) with $M$ replaced
by $N+M-1$ and in which $N-1$ of the $S$ fields have been suppressed. The
results of the analysis are of course invariant if we change the point
of view.

This paper is organized as follows. In section
{\bf 2} we present a brief review on the pseudo--differential analysis
of integrable systems with reference to the general pseudo--differential
operator (\ref{gpdo}) and write down the corresponding integrable hierarchy
(\ref{gh}). In
section {\bf 3} we will show that this general integrable hierarchy (\ref{gh})
admits a particular restriction which leads to the $(N, M)$--th hierarchy
(\ref{gkdvh}). As stated above,
the independent fields of (\ref{pdo}) satisfy two $(N+2M-1)$--dimensional
algebras. We refer in particular to the one corresponding to the \shs\, as the
$W(N, M)$--algebra. Its properties will be illustrated
in section {\bf 4}. In section {\bf 5} we will discuss the reduction
procedures. We will show that the $(N, M)$--th \kdvh\, (\ref{gkdvh})
possesses $M$ different reductions which are characterized by $M$
second class constraints $S_l=0 \, (l=1,2,\ldots,M)$. In particular, when we
impose the constraint $S_1=0$, the $(N, M)$--th \kdvh\, and $W(N, M)$
algebra are reduced to $(N+1, M-1)$--th \kdvh\, and $(N+1, M-1)$ algebra,
respectively. If we suppress all the $S_l$ fields in succession, we will
obtain the following two sequences
\a
(N, M)-{\rm th KdV}\stackrel{S_1=0}{\longrightarrow\longrightarrow}
(N+1, M-1)-{\rm th KdV}\stackrel{S_2=0}{\longrightarrow\longrightarrow}
(N+2, M-2)-{\rm th KdV}\ldots\ldots\0
\b\a
\stackrel{S_{M-1}=0}{\longrightarrow\longrightarrow}
(N+M-1, 1)-{\rm th KdV}\stackrel{S_M=0}{\longrightarrow\longrightarrow}
(N+M, 0)=(N+M)-{\rm th KdV}.\label{kdvred}
\b
for the hierarchies, and
\a
W(N, M)\stackrel{S_1=0}{\longrightarrow\longrightarrow}W(N+1, M-1)
\stackrel{S_2=0}{\longrightarrow\longrightarrow}W(N+2, M-2)\ldots\ldots\0
\b\a
\stackrel{S_{M-1}=0}{\longrightarrow\longrightarrow}
W(N+M-1, 1)\stackrel{S_M=0}{\longrightarrow\longrightarrow}
W(N+M, 0)=W_{N+M}.\label{algred}
\b
for the $W(N,M)$ algebras, respectively. The double arrow
${\longrightarrow\longrightarrow}$ means {\it reduces to} throughout
the paper.

Our results on reductions can be nicely summarized by means of a
Drinfeld--Sokholov representation, section {\bf 6}.

We also consider the dispersionless version of the $(N, M)$--th
\kdvh\, in section {\bf 7} and {\bf 8}. They constitutes the genus 0
part of the hierarchy in the context of matrix models, see \cite{bxII}.
We find that the dispersionless
$(N, M)$--th KdV hierarchy (\ref{disgkdvh}) admits, in addition to the ones
already considered, another type of reductions,
in which some $S_l$ fields are identified with one another instead
of being suppressed. In this way we
get more possible reductions. All of them can be seen as restrictions, i.e.
we can obtain the reduced hierarchies by imposing the constraints on the Lax
pair. Section {\bf9} contains some final remarks.

\section{The pseudo--differential analysis}

\setcounter{equation}{0}
\setcounter{subsection}{0}
\setcounter{footnote}{0}

In this section we give a short review of integrable differential
hierarchies, in particular we collect some results we will need later on
concerning pseudo--differential analysis, coadjoint orbit method and \bhs.
We refer to existing books and reviews (see \cite{dickey} and
references therein) for more detailed explanations.

\subsection{Pseudo-differential analysis}

Let us begin with the most general {\it pseudo--differential
operator}\, (denoted PDO)
\a
A=\sum_{-\infty}^{N} u_i(x)\partial_i.\label{gpdo}
\b
where $u_i(x)$'s are ordinary functions. As usual $\d=\ddx$ is the derivative
with respect to the space coordinate $x$, while $\d^{-1}$ is a formal
integration operator, which has the following properties
\a
&&\partial^{-1}\partial=\partial\partial^{-1}=1,\0\\
&&\partial^{-j-1}u=\sum_{v=0}^{\infty}(-1)^v{{j+v}\choose v}
u^{(v)}\partial^{-j-v-1},\label{glb}
\b
where $u'=\frac{\d u}{\d x}, u^{''}=\frac{\d^2 u}{\d x^2},\ldots,
u^{(v)}={{\d^v u}\over{\d x^v}}$. The above formula
together with the usual Leibnitz rule,
\a
\d u(x)=u(x)\d+u'(x),\qquad [\d, x]=1\0
\b
provides an algebraic structure on the whole operator space
formed by the general PDO's (\ref{gpdo}). We call this algebra
 {\it pseudo--differential algebra}\, $\wp$.

\def\pdo{pseudo--differential operator}
For any \pdo\, $A$ of type (\ref{gpdo}), we call order the number $N$
(the highest power of $\d$).  We will use the conventions
\a
\left\{\ba{l}
A_+:{\rm pure ~differential ~part~ of~ A;}\\
A_-:{\rm pure ~integration ~part~ of~ A;}\\\noal
A_{(i,j)}\equiv\sum_{l=i}^j u_l(x)\d^l,\qquad i>j;\\\noal
A_{(i)}\equiv u_i\d^i,\qquad -\infty<i\leq N.
\ea\right.\0
\b
Therefore any \pdo\, has a natural decomposition
\a
A=A_++A_-.\0
\b
which leads to an analogous decomposition in $\wp$
\a
\wp = \wp_+ +\wp_- \0
\b
In terms of this truncation, we can introduce the useful mapping
\cite{semenov}
\def\ro{{\cal R}}
\a
\ro A\equiv A_+-A_-.\label{ro}
\b
This mapping defines another algebraic structure on $\wp$
\a
[X, Y]_\ro\equiv \ha\Bigl([\ro X, Y]+[X, \ro Y]\Bigl)
=[X_+, Y_+]-[X_-, Y_-], \qquad\forall X, Y\in \wp.\label{rb}
\b
This $\ro$--commutator will be very important in our discussion on Hamiltonian
structures. But before that, we need more notations.
We call $u_{-1}(x)$ the residue of  pseudo--differential operator $A$ of
type (\ref{gpdo}) and denote it by
\def\res{\rm res}
\a
\res_{\d} A=u_{-1}(x)\qquad \hbox{ or}\qquad A_{(-1)},\0
\b
The following functional integral
\def\rtr{\rm Tr}
\a
<A>=\rtr(A)=\int (\res A)=\int u_{-1}(x)dx\label{ip}
\b
naturally defines an inner product on the algebra $\wp$, which
is nondegenerate, symmetric and invariant
\footnote{The product is symmetric since $<AB>=<BA>$, while invariance means
$<A[B,C]>=<[A,B]C>$. Occasionally we will also denote the product in other
ways: $<AB>=A(B)=Tr(AB)$. }.
Using this inner product, we may express a functional of the fields $u_i$'s
as follows
\a
f_X(A)=<AX>=A(X),\qquad X=\sum_{i}\d^i \chi_i(x)\0
\b
where $\chi_i(x)$'s are testing functions, and $X$ may be thought of as
a cotangent vector at the point $A$.
Generally, the cotangent vector $df$ is defined as
\a
\delta f(A)=f(A+\delta A)-f(A)\stackrel{\rm def}{=}<df, \delta A>.\0
\b
We denote by ${\cal F}(\wp)$  the functional space formed by
all functionals $f_X$ defined as above.

\subsection{Bi--Hamiltonian structure}

By means of the $\ro$--commutator introduced in the previous subsection, we may
define the adjoint action of the algebra $\wp$ on itself\footnote{
The usual adjoint action is ${\rm Ad}_Y X=[X, Y]$, here we use the same
notation to denote the adjoint action generated by the $\ro$--commutator.}
\a
{\rm Ad}_Y X\stackrel{\rm def}{=}[X, Y]_{\ro}.\0
\b
The coadjoint action of $\wp$ on the functional space ${\cal F}(\wp)$ is
specified by
\a
Ad^{*}_Yf_X(A)\stackrel{\rm def}{=}A(Ad_{-Y}X)=A([X,Y]_{\ro}),\0
\b
For a fixed $f_X$, as $Y$ varies in $\wp$, $Ad^{*}_Yf_X$
spans an orbit in the functional space ${\cal F}(\wp)$ which is called the
co--adjoint orbit. Since we may view $Y$ as a co--tangent vector, this
co--adjoint action naturally defines a Poisson structure on ${\cal F}(\wp)$
\a
\{f_X, f_Y\}_1(A)=A([X,Y]_\ro).\label{pb1}
\b
With respect to this Poisson bracket, the conserved quantities (Hamiltonians)
have very simple form
\a
H_r={{N}\over r}<A^{r\over{N}}>,\qquad \forall r\geq1;
\label{ghamiltonian}
\b
The corresponding cotangent vector at the point $A$ reads
\a
dH_r=A^{{r-N}\over{N}}_{(1-N, \infty)},\quad \forall r\geq1.\label{dhr}
\b
The time evolution of a function $f_X(A)$ is given by
\a
\ddt r f_X(A)=\{f_X(A), H_{r+N}\}_1
=<A, [X, dH_{r+N}]_\ro>,\0
\b
{}From eqs.(\ref{rb}, \ref{dhr}), we see that
\def\noal{\noalign{\vskip10pt}}
\a
&&~~<A, [X, dH_{r+N}]_\ro>=<A, [X_+, (A^{r\over{N}})_+]
   - [X_-, A^{r\over{N}}_{(1-N, -1)}]>\0\\ \noal
&&=-<[A, (A^{r\over{N}})_+], X_+>
  +<[A, A^{r\over{N}}_{(1-N, -1)}], X_->\0\\ \noal
&&=-<[A, (A^{r\over{N}})_+], X_+>
  +<[A, (A^{r\over{N}})_-], X_->\0\\ \noal
&&=<X, [(A^{r\over{N}})_+, A]>.\0
\b
In the second and the last equalities we have used the invariance of the
inner product; in the third step we have added a vanishing term
$<[A, A^{r\over{N}}_{(-\infty, -N)}], X_->$, since
\a
[A, A^{r\over{N}}_{(-\infty, -N)}]\in \wp_-,\quad{\rm and}\quad
<Y,Z>=0,\quad{\rm if}\quad Y,Z\in\wp_-.\0
\b
Therefore we have
\a
\ddt r f_X(A)=<X, [(A^{r\over{N}})_+, A]>,\label{ddtfxa}
\b
Suppose that $X$ is time independent; then we obtain the time evolution
equations
of the \pdo\, (\ref{gpdo})
\a
\ddt r A=[(A^{r\over{N}})_+, A], \qquad \forall r\geq1.\label{gh}
\b
This is the general integrable differential hierarchy we have obtained
from pseudo--differential analysis\footnote{If we define another
kind of $\ro$--operator such as
\a
\ro_l X=X_{l,\infty}-X_{-\infty, l-1},\0
\b
we obtain non--standard integrable hierarchies for some values of
$l$\cite{kuper}.}.
In order to prove its integrability
we derive the \shs, \cite{Magri}, compatible with the first, i.e.
a second Poisson bracket such that
\a
\{f_X, H_{r+N}\}_1=\{f_X, H_r\}_2.\label{gcompatibility}
\b
The LHS is just (\ref{ddtfxa}). Taking the residue of the following equality
\a
[\Bigl(A^{r\over{N}}\Bigl)_{(1-N,\infty)}, A]
=[A, \Bigl(A^{r\over{N}}\Bigl)_{(-\infty, -N)}],\0
\b
we get
\a
N\Bigl(A^{{r-N}\over{N}}_{(-N)}\Bigl)'
=[A^{r\over{N}}_{(1-N,\infty)}, A]_{(-1)}.\0
\b
On the other hand,
\a
&&~~(A^{r\over{N}})_+=(A\cdot A^{{r-N}\over{N}})_+
=\Bigl(A A^{{r-N}\over{N}}_{(1-N,\infty)}\Bigl)_+
 + \Bigl(A^{{r-N}\over{N}}\Bigl)_{(-N)}\0\\ \noal
&&=\Bigl((A^{{r-N}\over{N}})_{(1-N,\infty)}A\Bigl)_+
 + \Bigl(A^{{r-N}\over{N}}\Bigl)_{(-N)}.\0
\b
Therefore, we have
\a
&&~~~~{\rm LHS~of~eq.(\ref{gcompatibility})}
=<AX(A^{r\over{N}})_+>-<XA(A^{r\over{N}})_+>\0\\ \noal
&&=<AX\bigg(A\Bigl(A^{{r-N}\over{N}}\Bigl)_{(1-N,\infty)}\bigg)_+>
-<XA\Bigl(A^{{r-N}\over{N}}_{(1-N,\infty)}A\Bigl)_+>
+<(AX-XA)\Bigl(A^{{r-N}\over{N}}\Bigl)_{(-N)}>\0\\ \noal
&&=<AX\Bigl(A dH_r\Bigl)_+>-<XA\Bigl(dH_r A\Bigl)_+>
+{1\over{N}}\int [A, X]_{(-1)}\Bigl(\d^{-1}[dH_r, A]_{(-1)}\Bigl)\0\\ \noal
&&=<(XA)_+dH_rA>-<(AX)_+AdH_r>
+{1\over{N}}\int [A, dH_r]_{(-1)}\Bigl(\d^{-1}[A, X]_{(-1)}\Bigl)\0
\b
Now we can see that all the terms in the last equality are linear in the
cotangent vector $dH_r$ and in $X$. This is therefore a candidate for  a second
Poisson bracket. If we replace $dH_r$ by an arbitrary
cotangent vector $Y$, we have the general expression of  it (\ref{gh})
\a
\{f_X, f_Y\}_2(A)&=&<(XA)_+YA>-<(AX)_+AY>\0\\
&+&{1\over{N}}\int [A,Y]_{-1}\bigg(\d^{-1}[A,X]_{-1}\bigg).
\label{pb2}
\b
One can show that this Poisson bracket satisfies the Jacobi
identity \cite{os}, so it is indeed a well--defined Poisson structure.
With respect to this bracket, the conserved quantities are
the same as in eq.(\ref{ghamiltonian}), and generate
all the flows in the hierarchy (\ref{gh}).
Therefore we proved that the system (\ref{gh}) possesses a
bi--Hamiltonian structure and is therefore integrable.

\section{The $(N, M)$--th \kdvh}

\setcounter{equation}{0}
\setcounter{subsection}{0}
\setcounter{footnote}{0}

In the previous section we have constructed the \bhs\, for a general \pdo\,
and the corresponding integrable hierarchy.
In this section we will prove that
the general integrable hierarchy (\ref{gh}) admits a particular restriction
which leads to the $(N, M)$--th \kdvh\, (\ref{gkdvh}) with the Lax
operator (\ref{pdo}).

\subsection{The consistency of the general flows}

In order to show that the restricted Lax operator (\ref{pdo})  preserves
the hierarchy (\ref{gh}), and gives in fact the $(N, M)$--th \kdvh, we should
first prove the consistency of the flows defined by (\ref{gkdvh}),
and then check that they  commute.

For the usual $N$--th \kdvh\,(\kph), the scalar Lax operator is a pure
differential operator ( \pdo\,), so it is very easy to see its form
invariance under the time evolutions. For instance,
the scalar Lax operator involved in the $N$--th \kdvh\, is
\a
A_{\rm NKdV}=\d^N+\sum_{l=1}^{N-1}a_l\d^{N-l-1},\0
\b
Obviously we have
\a
\relax [\Bigl(A_{\rm NKdV}\Bigl)^{r\over N}_+, A_{\rm NKdV}]
=[ A_{\rm NKdV}, \Bigl(A_{\rm NKdV}\Bigl)^{r\over N}_-],\0
\b
The LHS is a purely differential operator, while the RHS indicates that this
operator
is of order $N-2$. This enables us to define the flows as follows
\a
\ddt r A_{\rm NKdV} =[\Bigl(A_{\rm NKdV}\Bigl)^{r\over N}_+, A_{\rm NKdV}].\0
\b
Therefore, the consistency of the flows in the $N$--th \kdvh\, (or \kph\, )
case is very simple.

However the present case is much more complicated since
the scalar Lax operator (\ref{pdo}) contains both differential and
pseudo--differential parts.  It is a nontrivial result that
the time evolutions of the Lax operator (\ref{pdo}) are
form invariant. This subsection is devoted to proving exactly this.

{\it Proposition 3.1} The LHS and the RHS of eq.(\ref{gkdvh}) are compatible.

{\it Proof.} To start with we write down the
$r$--th time evolution of the Lax operator (\ref{pdo})
\a
\ddt r L&=&\sum_{l=1}^{N-1}\Bigl({{\d a_l}\over{\d t_r}}\Bigl)\d^{N-l-1}
  +\sum_{l=1}^M\Bigl({{\d a_{N+l-1}}\over{\d t_r}}\Bigl)
{1\over{\d-S_l}}{1\over{\d-S_{l-1}}}\ldots{1\over{\d-S_1}}\0\\
&+&\sum_{l=1}^M\sum_{k=1}^l
a_{N+l-1}{1\over{\d-S_l}}\ldots{1\over{\d-S_k}}\Bigl({{\d S_k}\over{\d t_r}}
\Bigl){1\over{\d-S_k}}\ldots{1\over{\d-S_1}}.\label{ddtrl}
\b
The problem is to prove that the LHS of eq.(\ref{gkdvh}) contains the same type
of terms. Our first task will thus be to classify the terms in RHS of
this equation. To this end we introduce a new basis set
for the pseudo--differential operator algebra $\wp$. We split the basis
set into several classes:

{\it the differential operator class}: we adopt the usual basis
\a
\{f_i\d^i, i\in {\bf Z}_+\},\0
\b
with arbitrary functions $f_i$'s. By definition this basis spans the
operator space $F^0$.

{\it the first class of integration operators}:
\a
f_l{1\over{\d-S_l}}{1\over{\d-S_{l-1}}}\ldots{1\over{\d-S_1}},
 \qquad 1\leq l\leq M;\0
\b
where $f_l$'s are arbitrary functions.  For a fixed $l$, the above operators
span  by definition the space $F^1_l$. We also define the direct sum
space $F^1=\oplus_{l}F^1_l$.

{\it the second class of integration operators}:
\a
f_{lk}{1\over{\d-S_l}}\ldots{1\over{\d-S_k}}g
{1\over{\d-S_k}}\ldots{1\over{\d-S_1}},\quad
1\leq l\leq M; \quad
1\leq k\leq l.\0
\b
Once again $f_{lk}$'s are arbitrary functions. We denote by $F^2_{l,k}$
the space spanned by the above operators with fixed $l,k$, and
\a
F^2=\{\oplus_{l,k}F^2_{l,k},\quad 1\leq l\leq M; \quad 1\leq k\leq l\}.\0
\b
All these operators  are linearly independent. Actually they do not
constitute a complete basis of the algebra $\wp$. However, since the
residual basis subset will not show up in our later discussion, we do not
need to write them out explicitly. Going back to
eq.(\ref{ddtrl}), we immediately see that its RHS contains all these terms.
We will say that a pseudodifferential operator takes the standard form
$F^0, F^1$ or $F^2$ if it belongs to the corresponding vector subspace.

What we should do next is to prove that the RHS of eq.(\ref{gkdvh}) can
be recast into these standard forms.

In order to simplify  the calculation of the commutator in eq.(\ref{gkdvh}),
we adopt the notation \def\opo{{\cal O}}
\a
\opo_{(r)}\stackrel{\rm def}{=}\Bigl(L^{r\over{N}}\Bigl)_+
=\d^r+\sum_{l=1}^{r-2}\alpha_l\d^l, \qquad \forall r\geq1.\0
\b
where the $\alpha_l$'s are certain functions of the $\{a_l\}$'s and the
$\{S_l\}$'s. The commutator in eq.(\ref{gkdvh}) can be decomposed into
two parts
\a
[\opo_{(r)}, L]=[\opo_{(r)}, L_+]+[\opo_{(r)}, L_-],\0
\b
Obviously the first part is a purely differential operator of order $N-1$,
while the second part can be further simplified
\a
&&~~[\opo_{(r)}, L_-]=[\opo_{(r)},
  \sum_{l=1}^M a_{N+l-1}{1\over{\d-S_l}}{1\over{\d-S_{l-1}}}
  \ldots {1\over{\d-S_1}}]\0\\
&&=\sum_{l=1}^M [\opo_{(r)}, a_{N+l-1}]
{1\over{\d-S_l}}{1\over{\d-S_{l-1}}}
  \ldots {1\over{\d-S_1}}\0\\
&&-\sum_{l=1}^M\sum_{k=1}^l a_{N+l-1}{1\over{\d-S_l}}\ldots{1\over{\d-S_k}}
  [\opo_{(r)}, \d-S_k]{1\over{\d-S_k}}\ldots {1\over{\d-S_1}},\label{prl-}
\b
Here we have used
\a
[\opo_{(r)}, {1\over{\d-S_k}}]=-{1\over{\d-S_k}}
[\opo_{(r)}, \d-S_k]{1\over{\d-S_k}}.\0
\b
The remaining two commutators in the expression (\ref{prl-})
only involve purely differential operators, both of them are
of order $r-1$. Therefore we find that the RHS
of eq.(\ref{gkdvh}) contains the following three kinds of terms
\def\il{{\bf I}_l}
\def\jlk{{\bf J}_{l,k}^{(r-1)}}
\a
\left\{\ba{ll}
G_0:&{\rm purely~differential~operator~part}\qquad [\opo_{(r)}, L_+]\\ \noal
\il^{(r)}:&\opo_{(r-1)}{1\over{\d-S_l}}{1\over{\d-S_{l-1}}}
  \ldots {1\over{\d-S_1}}, \qquad 1\leq l\leq M;\\ \noal
\jlk:&\opo_{(0)}{1\over{\d-S_l}}\ldots{1\over{\d-S_k}}
  \opo_{(r-1)}{1\over{\d-S_k}}\ldots {1\over{\d-S_1}},\quad
  1\leq k\leq l\leq M.
\ea\right.\0
\b
where $\opo_{(r)}$ is  an arbitrary {\it purely} differential operator
of $r$--th order, while $\opo_{(0)}$ is an ordinary function.
We will denote
\a
&&G_1=\{\sum_{l}\il, \qquad 1\leq l\leq M\},\0\\
&&G_2=\{\sum_{l,k}\jlk, \qquad 1\leq k\leq l\leq M\}.\0
\b
We will need one more operator space, $G^3$,
which contains the following type of terms
\def\vlk{{\bf V}_{l,k}^{(r-1)}}
\a
\vlk:\quad \opo_{(0)}{1\over{\d-S_l}}\ldots{1\over{\d-S_{k+1}}}
  \opo_{(r-1)}{1\over{\d-S_k}}\ldots {1\over{\d-S_1}},\quad
  1\leq k\leq l\leq M.\0
\b
Here $\opo_{(r-1)}$ is another purely differential operator.
In the case $k=l$, this is the same as $G_1$, i.e.
\a
{\bf V}_{l,l}^{(r)}={\bf I}^{(r)}_l.\label{vgotoi}
\b
In the remaining part of this subsection we will show how to recast
the above terms into the standard forms $F^0, F^1, F^2$.

\vskip0.4cm
\noindent
\underline{\bf $G_0$--terms:}
\vskip 0.3cm

We notice that
\a
[\Bigl(L^{r\over{N}}\Bigl)_+, L]=[L, \Bigl(L^{r\over{N}}\Bigl)_-],\0
\b
so the power expansion of the above commutator has order  $<N-1$,
i.e. $G_0$--terms are pure differential operators of $(N-2)$--th order, i.e.
they
have exactly the same form as  $F^0$--terms. Therefore $G_0$--terms
contribute to the time evolutions of the fields $a_l~(1\leq l\leq N)$.

\vskip0.4cm
\noindent
\underline{\bf $G_1$--terms:}
\vskip 0.3cm

In order to simplify $G_1$--terms, we recall that for any purely differential
operator $\opo_{(r)}$ of $r$--th order, and any ordinary function $f$, we can
find two $(r-1)$--th order purely differential operators $\opo_{(r-1)}$
and ${\tilde \opo}_{(r-1)}$ such that
\a
\opo_{(r)}=(\d-f)\opo_{(r-1)}+g,\qquad {\rm and}\qquad
\opo_{(r)}={\tilde \opo}_{(r-1)}(\d-f)+{\tilde g}.\label{rtor-1}
\b
where $g$ and ${\tilde g}$ are two ordinary functions specified by the above
equalities. The second equality immediately leads to the following results
\a
\il^{(r-1)}\Longrightarrow \il^{(r-2)}\oplus F_l^1.\0
\b
repeating this procedure we can reduce all   the $G_1$--terms to $F^1$
(possibly
$F^0$) terms. For instance
\a
&&~~[\Bigl(L^{r\over{N}}\Bigl)_+, a_{N+l-1}]
  =\opo_{(r-1)}=f_0+\opo_{(r-2)}(\d-S_l)\0\\
&&=f_0+\Bigl(f_1+\opo_{(r-3)}(\d-S_{l-1})\Bigl)(\d-S_l)\label{g1example}\\
&&=\qquad \ldots\ldots\0\\
&&=f_0+\sum_{i=1}^{l-1}f_i(\d-S_{l-i+1})(\d-S_{l-i})\ldots(\d-S_l)\0\\
&&+\opo_{(r-l-1)}(\d-S_1)(\d-S_2)\ldots(\d-S_l),\0
\b
where all the functions $f_l$'s and the operator $\opo_{(r-l-1)}$ are
completely determined by the commutator. Therefore
the $G_1$--term
\a
[(L^{r\over{N}})_+, a_{N+l-1}]{1\over{\d-S_l}}{1\over{\d-S_{l-1}}}
\ldots{1\over{\d-S_1}}\0
\b
becomes an $F^1$--term. If $r\ge l+1$, the last
term in eq.(\ref{g1example}) contributes an $F^0$--term $\opo_{(r-l-1)}$.
We may summarize the procedure with the following diagram
\a
\left.\ba{ccccccccccc}
\il^{(r)}&\longrightarrow&
{\bf I}_{l-1}^{(r-1)}&\longrightarrow&
{\bf I}_{l-2}^{(r-2)}&\longrightarrow&
\ldots&\longrightarrow&
{\bf I}_1^{(r-l-1)}&\longrightarrow&
{\bf I}_0^{(r-l)}\\
\downarrow&&\downarrow&&\downarrow&&&&\downarrow&&\Vert\\
F^1_l&&F^1_{l-1}&&F^1_{l-2}&&&&F^1_1&&F^0
\ea\right.\label{igotof1}
\b

\vskip0.4cm
\noindent
\underline{\bf $G_2$--terms:}
\vskip 0.3cm

Now let us consider the $G_2$--terms. Their form is
\a
&&\jlk=a_{N+l-1}{1\over{\d-S_l}}\ldots{1\over{\d-S_k}}
  [(L^{r\over{N}})_+, \d-S_k]{1\over{\d-S_k}}\ldots{1\over{\d-S_1}}\0\\\noal
&&=a_{N+l-1}{1\over{\d-S_l}}\ldots{1\over{\d-S_k}}\opo_{(r-1)}
  {1\over{\d-S_k}}\ldots{1\over{\d-S_1}}\0\\\noal
&&=a_{N+l-1}{1\over{\d-S_l}}\ldots{1\over{\d-S_k}}\opo_{(0)}
  {1\over{\d-S_k}}\ldots{1\over{\d-S_1}}\0\\\noal
&&+a_{N+l-1}{1\over{\d-S_l}}\ldots{1\over{\d-S_{k+1}}}\opo_{(r-2)}
  {1\over{\d-S_k}}\ldots{1\over{\d-S_1}},\0
\b
in the last step we have used the first equality in eqs.(\ref{rtor-1}).
This decomposition simply means
\a
\jlk\Longrightarrow {\bf V}^{(r-2)}_{l,k}\oplus F^2_{l,k}.\label{jgotov}
\b
Here and in the following $\Longrightarrow$ means decomposition.
In other words, $G_2$--terms can be decomposed into $F^2$ and $G_3$ terms.
Therefore it remains for us to treat $G_3$--terms.

\vskip0.4cm
\noindent
\underline{\bf $G_3$--terms:}
\vskip 0.3cm

As we know, a $G_3$--term is of the form
\a
\vlk=a_{N+l-1}{1\over{\d-S_l}}\ldots{1\over{\d-S_{k+1}}}\opo_{(r-2)}
  {1\over{\d-S_k}}\ldots{1\over{\d-S_1}},\0
\b
With a simple calculation we get
\a
&&~~~{1\over{\d-S_{k+1}}}\opo_{(r-2)}
=\opo_{(r-2)}{1\over{\d-S_{k+1}}}+[{1\over{\d-S_{k+1}}}, \opo_{(r-2)}]\0\\\noal
&&=\opo_{(r-2)}{1\over{\d-S_{k+1}}}-{1\over{\d-S_{k+1}}}
[\d-S_{k+1}, \opo_{(r-2)}]{1\over{\d-S_{k+1}}}\0\\\noal
&&=\opo_{(r-2)}{1\over{\d-S_{k+1}}}-{1\over{\d-S_{k+1}}}
-{1\over{\d-S_{k+1}}}\opo_{(r-3)}{1\over{\d-S_{k+1}}}.\0
\b
This result shows
\a
\vlk\Longrightarrow {\bf V}^{(r-2)}_{l, k+1}
\oplus {\bf J}^{(r-1)}_{l,k+1},\label{vgotoj}
\b
The crucial feature of this step is that we have moved the operators in
$G_3$ one step to the right.

Combining the procedures (\ref{jgotov}) and (\ref{vgotoj}), and remembering
the fact (\ref{vgotoi}), we can recast the $G_2, G_3$--terms into the standard
forms. Diagrammatically
\a
\left.\ba{ccccccccccc}
{\bf V}_{l,k}^{(r-1)}&\Longrightarrow&
{\bf V}_{l,k+1}^{(r-2)}&\Longrightarrow&
{\bf V}_{l,k+2}^{(r-3)}&\Longrightarrow&\ldots&\Longrightarrow&
{\bf V}_{l,l}^{(r-l+k-1)}&=&{\bf I}_l^{(r-l+k-1)}\\
\uparrow&\searrow&\uparrow&\searrow&\uparrow&\searrow&\ldots&
\searrow&\uparrow&&\\
{\bf J}_{l,k}^{(r)}& &{\bf J}_{l,k+1}^{(r-1)}&&{\bf J}_{l,k+2}^{(r-2)}&&
\ldots&&{\bf J}_{l,l}^{(r-l+k)}&&\\
\downarrow&&\downarrow&&\downarrow&&\ldots&&\downarrow&&\\
F^2_{l,k}&&F^2_{l,k+1}&&F^2_{l,k+2}&&\ldots&&F^2_{l,l}&&
\ea\right.\label{jvgotoif2}
\b
The last term in the first line can be treated through the procedure
(\ref{igotof1}).
So we finally proved that the RHS of eq.(\ref{gkdvh}) has exactly
the same form as its LHS. Comparing their explicit expressions, we
can obtain the equations of motion of the fundamental fields.

\subsection{Commutativity of the flows}

Our next problem is to prove the following

{\it Proposition 3.2} All the flows defined in (\ref{gkdvh})
commute with one another, i.e.
\a
\ddt l\Bigl(\ddt r L\Bigl)=\ddt r\Bigl(\ddt l L\Bigl).
\b

{\it Proof.} This is elementary to prove since we have explicitly
\a
{\rm RHS}&=&[[L^{{r\over N}}_+,L^{{l\over N}}]_+, L]
+[L^{{l\over N}}_+, [L^{{r\over N}}_+, L]]\0\\
&=&-[[L^{{r\over N}}_-, L^{{l\over N}}_+]_+, L]
-[L^{{r\over N}}_+, [L, L^{{l\over N}}_+]]
-[L, [L^{{l\over N}}_+, L^{{r\over N}}_+]]\0\\
&=&[[L^{{l\over N}}_+,L^{{r\over N}}]_+, L]
+[L^{{r\over N}}_+, [L^{{l\over N}}_+, L]]={\rm LHS}.\0
\b
In the second equality we have used the Jacobi identity.

\subsection{The non--linear evolution equations}

Now let us give here the first non--trivial flow of the hierarchy (\ref{gkdv})
(see Appendix {\bf A} for the derivation),
\ai
&&\ddt 2 a_l=a_l^{''}+2a'_{l+1}-{2\over{N}}
  \left(\ba{c}N\\l+1\ea\right)a_1^{(l+1)}\0\\\noal
&&\qquad\qquad-{2\over{N}}\sum_{k=1}^{l-1}
  \left(\ba{c}N-k-1\\l-k\ea\right)a_ka_1^{(l-k)},
   \qquad 1\leq l\leq N-1;\label{gkdv}\\\noal
&&\ddt 2 a_{N+l}=a^{''}_{N+l}+2a'_{N+l+1}
  +2a'_{N+l}S_{l+1}+2a_{N+l}\Bigl(\sum_{k=1}^{l+1} S_k\Bigl)',
  ~0\leq l\leq M-1;\label{pkdv}\\
&&\ddt 2 S_l={2\over{N}}a_1'+2S_lS_l'-S_l^{''}
  -2\Bigl(\sum_{k=1}^{l-1} S_k\Bigl)^{''},\qquad   1\leq l\leq M.
\label{ppkdv}
\bj
This set of non--linear evolution equations are crucial
objects, all the important ingredients like the \bhs\ and the algebraic
structures,
as well as the conserved quantities, are rooted in these equations; the
pseudo--differential analysis is only a powerful tool to make them explicit.
In other words,
we can say that the above equations define a bi--Hamiltonian system with
two Poisson structures  (\ref{pb1}) and (\ref{pb2}),  the hierarchical
equations (\ref{gkdvh}) can be thought of as symmetries of these non--linear
equations.

We end this section with a few remarks.

\noindent
$(i)$. Comparing the first $N-1$ equations (\ref{gkdv}) with
the $N$--th KdV equations, we see that they are
the same, except that the
$a_{N}$ field is involved in the time evolution of the field $a_{N-1}$
in eq.(\ref{gkdv}). Therefore,
if we set $a_{N+l-1}=S_l=0, l=1,2,\ldots,M$; we recover the $N$--th
\kdvh.
So we may view this set of equations (\ref{gkdv})---(\ref{ppkdv})
as a generalization or a perturbation of the $N$--th KdV equations
by means of the fields $\{a_{N+l-1}, S_l;~ 1\leq l\leq M\}$).

\noindent
$(ii)$. In the $N=1$ case, the integrable hierarchy
(\ref{gkdvh}) is nothing but the $2M$--field representation of
\kph\,\cite{bx3}.

In this sense we can say that the $(N, M)$--th \kdvh\, (\ref{gkdvh})
contains both
$N$--th \kdvh\, and the multi--boson representations of \kph.

\section{$W(N, M)$--algebra (or the extended $W$--algebra)}

\setcounter{equation}{0}
\setcounter{subsection}{0}
\setcounter{footnote}{0}

As we know, the \shs\, of the $N$--th \kdvh\, gives rise to a $W_N$
algebra, and the \shs\, of the \kph\, leads to a $W_\infty$ algebra. We will
see
that the \bhs\, of the integrable hierarchy (\ref{gkdvh}) results in two
{\it finite}\, dimensional algebras, which generate the extended
$W_{1+\infty}$ and $W_\infty$ algebras, respectively.

Before we proceed we wish to clarify the meaning of our choice of coordinates
in (\ref{pdo}). Up to now in fact we have only used
$\{a_l,l=1,2,\ldots,N-1; a_{N+l-1}, S_l,l=1,2,\ldots, M\}$
as our dynamical fields ({\it coordinates}\,).
However, one might choose any other system of coordinates. For instance, we can
introduce a new set of coordinates in the following way
\a
(\ln r_l)'=-S_l,\qquad q_l={a_{N+l-1}\over{r_l}},\qquad
 1\leq l\leq M;\label{astoqr}
\b
Noting
\a
{1\over{\d}}r_l=r_l{1\over{\d+(\ln r_l)'}},\0
\b
we can immediately rewrite (\ref{pdo}) as follows
\a
L=\d^{N}+\sum_{l=1}^{N-1} a_l\d^{N-l-1}
 +\sum_{l=1}^M q_l\d^{-1}\Bigl({r_l\over r_{l-1}}\Bigl)\d^{-1}\ldots
\Bigl({r_2\over r_{1}}\Bigl)\d^{-1}r_1.\label{pdoqr}
\b
Each set of coordinates have their own advantages (and in general
different physical meaning). For
example, the first choice (\ref{pdo}) leads to {\it simple}
and {\it local} Poisson algebras, while the Poisson algebras in the second
choice (\ref{pdoqr}) will contain non--local terms. We can think of the
passage from one set of coordinates to another set as a field--dependent
gauge transformation of the corresponding linear system (see the
representations of the Lax operators in terms of
linear systems of section {\bf 6}).
For this reason we will refer to different choices of the coordinates as
``different gauges".

We notice however that we can write
\a
L \stackrel{\rm def}{=}\d^{N}+\sum_{l=1}^{N-1} a_l\d^{N-l-1}
 +\sum_{n=0}^{\infty} u_n\d^{-n-1},\label{ul}
\b
where
\a
&&u_n=\sum_{l=1}^{n}\sum_{\sum_{i=1}^l\mu_i=n-l+1}
a_{N+l-1}(-\d+S_l)^{\mu_l}(-\d+S_{l-1})^{\mu_{l-1}}
\ldots(-\d+S_1)^{\mu_1}\cdot1,\0\\\noal
&&\qquad\qquad\qquad\qquad 0\leq n\leq M-1;\0\\\noal
&&u_n=\sum_{l=1}^{M}\sum_{\sum_{i=1}^l\mu_i=n-l+1}
a_{N+l-1}(-\d+S_l)^{\mu_l}(-\d+S_{l-1})^{\mu_{l-1}}
\ldots(-\d+S_1)^{\mu_1}\cdot1,\label{ulas}\\\noal
&&\qquad \qquad\qquad \qquad M\leq n\leq \infty.\0
\b
The {\it form} of the operator $L$ is gauge invariant and by (a suggestive)
abuse of language we call
the coordinates $\{a_l, 1\leq l\leq N-1; u_l, l\geq0\}$ `gauge invariant'.

The expressions (\ref{ulas}) have the following remarkable features:
\begin{itemize}
\item
the $u_l$'s are linear in the $a_l~(1\leq l\leq N+M-1)$ fields, but
highly non--linear in the $S_l$ fields;
\item
the $u_l$'s do not contain any derivative of the
$a_l~(1\leq l\leq N+M-1)$ fields
while they contain derivatives of the $S_l$ fields.
\item
these formulas show that
the subsets of fields $\{a_l, 1\leq l\leq N-1\}$  and
$\{a_{N+l}, 0\leq l\leq M-1\}$ introduced in (\ref{pdo}) are of
quite different nature, since the former set is canonical
the other is not. However for later convenience we will keep the notation
$\{a_l\}$ for both subsets.

\item in the gauge (\ref{pdoqr}), the $u_l$'s will be linear in the $q_l$
fields, but non--linear and non--polynomial in the $r_l$ fields.

\end{itemize}

In the following we will mostly work with the gauge (\ref{pdo}).

\subsection{The extended $W_\infty$ algebras}

If we substitute the Lax operator (\ref{ul}) into (\ref{pb1}) and (\ref{pb2}),
we will get two {\it infinite}\, dimensional algebras, which we will call
extended $W_{\infty}$ algebras. The calculation is in principle
straightforward, but it requires some skillful use of suitable
techniques to simplify the results. However we will omit any detail and only
give the results.

\vskip3mm
\noindent
\underline{\bf The extended $W_{1+\infty}$ algebra:}
\vskip2mm

{\it Proposition 4.1} The first Poisson structure (\ref{pb1}) leads to the
following explicit algebra
\def\dxy{\delta(x-y)}
\ai
\{a_i, a_j\}_1&=&\bigg[\bigg(\left(\ba{c} j\\ N-i-1\ea\right)
  +(-1)^{i+j-N}\left(\ba{c} i\\ N-j-1\ea\right)\bigg)\d^{i+j-N+1}\0\\
&+&\sum_{l=i}^{N-1}(-1)^{i+j-l-N-1}\left(\ba{c} i-l-1\\ N-j-1\ea\right)
  \d^{i+j-l-N}a_l\label{aiaj1}\\
&+&\sum_{l=j}^{N-1}\left(\ba{c} j-l-1\\ N-i-1\ea\right)
  a_l\d^{i+j-l-N}\bigg]\dxy\0\\
\{a_i, u_j\}_1&=&0,\label{aiuj1}\\
\{u_i, u_j\}_1&=&\Bigl[\sum_{l=1}^j\left(\ba{c} j\\ l\ea\right)
  \d^l u_{i+j-l}+\sum_{l=1}^i(-1)^{l+1}\left(\ba{c} i\\ l\ea\right)
  u_{i+j-l}\d^l\Bigl]\dxy.\label{uiuj1}
\bj
In the above Poisson brackets (as well as in subsequent ones)
we use a shorthand notation. They must be
understood in the following way
\a
\{f(x), g(y)\}_1={\hat O}(x)\dxy,\qquad \forall f,g;\0
\b
Moreover either the indices of $a_l$ ($u_l$) fields are in the region
$(1\leq l\leq N-1)$ ($l\geq0$) or the corresponding terms are
understood to be absent. Similarly, either the powers of $\d$ are non--negative
or the corresponding terms are absent. We remark that a central term is
contained
only in eq.(\ref{aiaj1})

The above Poisson algebra is a direct sum of two sub--algebras,
the sub--algebra (\ref{aiaj1}) coincides
exactly with the first Poisson algebra in the $N$--th
\kdvh. From eq.(\ref{uiuj1}), we see that
\a
\{u_1, u_1\}_1=(u_1\d+\d u_1)\dxy,\0
\b
Therefore $u_1$ can be viewed as conformal tensor of weight 2.
The Poisson bracket $\{u_1, u_0\}_1$ indicates that
$u_0$ has conformal spin 1. Such sub--algebra (\ref{uiuj1})
is commonly referred to as a $W_{1+\infty}$ algebra.

\vskip3mm
\noindent
\underline{\bf The extended $W_{\infty}$ algebra:}
\vskip2mm

The second Poisson algebra is much more complicated than the former one.

{\it Proposition 4.2} The explicit form of the second Poisson brackets is
\ai
&&\{a_i, a_j\}_2=\bigg[c_{ij}\d^{i+j+1}
+\bigg(\sum_{l=i+j-N+1}^i(-1)^{l+1}\left(\ba{c}i\\ l\ea\right)+
    \sum_{l=j+1}^{i+j-1} b^1_{ijl}\bigg)a_{i+j-l}\d^l\0\\\noal
&&  +\sum_{l=1}^{i+j-N}\bigg(\left(\ba{c} j\\ l\ea\right)\d^l u_{i+j-l-N}
  +(-1)^{l+1}\left(\ba{c} j\\ l\ea\right) u_{i+j-l-N}\d^l\bigg)\0\\\noal
&&+\bigg(\frac{1}{N}\sum_{l=i+1}^{i+j-1}(-1)^{i+l+1}
   \left(\ba{c}N\\ i+1\ea\right)\left(\ba{c}N+l-i-j-1\\ l-i\ea\right)
   +\sum_{l=i+j-N+1}^{i+j-1}b^2_{ijl}\bigg)\d^l a_{i+j-l}
     \label{aiaj2}\\\noal
&&+\sum_{l=1}^{i-1}\bigg(\sum_{k=1}^{i-l-1}(-1)^{k+1}
   \left(\ba{c}i-l-1\\ k\ea\right)a_{i+j-l-k-1}\d^k a_l
   +\sum_{k=i+j-l-N}^{i+j-l-2}b^3_{ijlk}a_l\d^k a_{i+j-l-k-1}\bigg)\0\\\noal
&&+\sum_{l=1}^{i-1}\sum_{k=1}^{i+j-l-N-1}\bigg((-1)^{k+1}
   \left(\ba{c}i-l-1\\ k\ea\right)u_{i+j-l-k-N-1}\d^k a_l
   +\left(\ba{c}j-l-1\\ k\ea\right)a_l\d^k u_{i+j-l-k-N-1}\bigg)\0\\\noal
&&+\frac{1}{N}\sum_{l=1}^{i-1}\sum_{k=1}^{j-1}(-1)^{j+k+1}
   \left(\ba{c}N-l-1\\ i-l\ea\right)
   \left(\ba{c}N-k-1\\ j-k\ea\right)a_l\d^{i+j-l-k-1}a_k\bigg]\dxy;\0\\\noal
&&\{a_i, u_j\}_2=\bigg[\sum_{l=1}^{N+j}\left(\ba{c} N+j\\ l\ea\right)
   \d^l u_{i+j-l}+ \sum_{l=1}^{i}(-1)^{l+1}\left(\ba{c} i\\ l\ea\right)
   u_{i+j-l}\d^l\0\\\noal
&&+\sum_{l=1}^{N-1}\sum_{k=1}^{N+j-l-1}\left(\ba{c} N+j-l-1\\ k\ea\right)
  a_l\d^k u_{i+j-l-k-1}
-{1\over{N}}\sum_{l=0}^{j-1}\left(\ba{c}N\\ i+1\ea\right)
  \left(\ba{c} j\\ l\ea\right)\d^{i+j-l} u_l\0\\\noal
&&+\sum_{l=1}^{i-1}\sum_{k=1}^{i-l-1}(-1)^{k+1}
  \left(\ba{c}i-l-1\\ k\ea\right)u_{i+j-l-k-1}\d^k a_l\label{aiuj2}\\\noal
&&-{1\over{N}}\sum_{l=1}^{i-1}\sum_{k=0}^{j-1}
  \left(\ba{c}N-l-1\\ i-l\ea\right)
  \left(\ba{c} j\\ k\ea\right)a_l\d^{i+j-l-k-1} u_k\bigg]\dxy;\0\\\noal
&&\{u_i, u_j\}_2=\bigg[\sum_{l=1}^{N+j}\left(\ba{c} N+j\\ l\ea\right)
  \d^l u_{i+j+N-l}+\sum_{l=1}^{N+i}(-1)^{l+1}\left(\ba{c}
  N+i\\ l\ea\right)u_{i+j+N-l}\d^l\0\\\noal
&&+\sum_{l=0}^{i-1}\bigg(\sum_{k=1}^{j-l-1}\left(\ba{c} j-l-1\\ k\ea\right)
  u_l\d^k u_{i+j-l-k-1}+\sum_{k=1}^{i-l-1}(-1)^{k+1}
\left(\ba{c}i-l-1\\ k\ea\right)u_{i+j+N-l-k-2}\d^k u_l\bigg)\0\\\noal
&&+\sum_{l=1}^{N-1}\sum_{k=1}^{N+j-l-1}\left(\ba{c} N+j-l-1\\ k\ea\right)
  a_l\d^k u_{i+j+N-l-k-1}\label{uiuj2}\\\noal
&&+\sum_{l=1}^{N-1}\sum_{k=1}^{N+i-l-1}(-1)^{k+1}
  \left(\ba{c} N+i-l-1\\ k\ea\right)u_{i+j+N-l-k-1}\d^k a_l\0\\\noal
&&+{1\over{N}}\sum_{l=1}^i\sum_{k=1}^j(-1)^{l+1}
   \left(\ba{c}i\\ l\ea\right)\left(\ba{c} j\\ k\ea\right)
   u_{i-l}\d^{l+k-1}u_{j-k}\bigg]\dxy.\0
\bj
where
\a
&&c_{ij}=\frac{(-1)^j}{N}\left(\ba{c}N\\ i+1\ea\right)
         +\sum_{l=0}^i(-1)^{i+l+1}\left(\ba{c}N\\ l\ea\right)
          \left(\ba{c}N+i-l\\ N-j\ea\right)\0\\\noal
&&b^1_{ijl}=\frac{(-1)^j}{N}\left(\ba{c}N\\ j+1\ea\right)
          \left(\ba{c}N+l-i-j-1\\ l-j\ea\right)\0\\\noal
&&+\sum_{k=0}^{l-j-1}(-1)^{k+l}\left(\ba{c}N+l-i-j-1\\ k\ea\right)
          \left(\ba{c}N+l-j-k-1\\ N-j-1\ea\right)\0\\\noal
&&b^2_{ijl}=\sum_{k=0}^i(-1)^{l+k}\left(\ba{c}N\\ k\ea\right)
          \left(\ba{c}N+l-j-k-1\\ N-j-1\ea\right)\0\\\noal
&&b^3_{ijlk}=\sum_{r=0}^{i-l-1}(-1)^{r+k}\left(\ba{c}N-l-1\\ r\ea\right)
          \left(\ba{c}N+k-i-r-1\\ N-j-1\ea\right).\0
\b

{}From eq.(\ref{aiaj2}) can immediately extract a Virasoro subalgebra
\a
\{a_1, a_1\}_2=\bigg(\ha\left(\ba{c}N+1\\3\ea\right)\d^3+a_1\d+\d
  a_1\bigg)\dxy,\label{vir}
\b
i.e. $a_1$ can be interpreted as a semi--classical energy momentum tensor.
The Poisson brackets between $a_1$ and the other fields tell us the conformal
dimensions (or spin contents) $[~\cdot~]$ of our coordinates
\a
&&[a_l]_{\rm conf}=l+1,\qquad l=1,2,\ldots, N+M-1;\0\\
&&[S_l]_{\rm conf}=1, \qquad\quad l=1,2,\ldots, M;\0\\
&&[u_l]_{\rm conf}=N+l+1, \qquad\quad l=0,1,\ldots,\infty.\0
\b
If we assign to $\delta$--function a conformal weight $1$, then we
have
\a
[\{, \}_2]_{\rm conf}=0.\0
\b
The algebra (\ref{aiaj2}--\ref{uiuj2}) contains fields with spins from
$2$ to infinity, and is linear or bilinear in the gauge invariant functions.
There exist central extensions represented by the coefficients $c_{ij}$.
For this reason we call this algebra the extended $W_\infty$ algebra.

\subsection{The finite dimensional algebras associated to the $(N, 1)$--th
 hierarchy}

The above algebras are independent of the particular coordinatization
we choose for the Lax operator. However, the physical meaning of a
hierarchy may essentially depend on the gauge, in particular on the number
of fields. Therefore we are very much interested in the algebras formed by
the coordinates we choose. In particular the independent fundamental
coordinate fields in the integrable hierarchy (\ref{gkdvh}) are finite.
Therefore we expect two finite algebras corresponding to the two
compatible Poisson structures. They in turn generate the infinite
dimensional algebras of the previous subsection. We can derive
these algebras from the infinite dimensional algebras
(\ref{aiaj1}--\ref{uiuj1})
and (\ref{aiaj2}--\ref{uiuj2}) by making use of the expressions (\ref{ulas}).
As a example, we consider the $(N, 1)$--th hierarchy, in which
the scalar Lax operator is
\a
L=\d^{N}+\sum_{l=1}^{N-1} a_l\d^{N-l-1}+a_{N}\frac{1}{\d-S_1}.\label{ln1}
\b
The gauge invariant functions are $\{a_1,a_2,\ldots,a_{N-1}\}$ and a set
of infinite many functions( {\it they are generated by only two fields}\, )
\a
u_l=a_{N}\alpha_l,\qquad\alpha_l\equiv(-\d+S_1)^l\cdot1,
\qquad l\geq0.\label{ulas1}
\b

{\it Proposition 4.3}  The first Hamiltonian structure leads to the following
Poisson algebra
\ai
&&\{a_i, a_j\}_1={\rm (\ref{aiaj1})},\qquad 1\leq i,j\leq N-1;\\
&&\{a_i, a_{N}\}_1=0,\qquad i=1,2,\ldots,N;\\
&&\{a_{i}, S_1\}_1=\delta_{i,N}\ddxy,\qquad i=1,2,\ldots,N;\\
&&\{S_1, S_1\}_1=0,
\bj
This $(N+1)$ dimensional algebra generates  the $W_{1+\infty}$ algebra
(\ref{aiaj1}--\ref{uiuj1}) through the transformation (\ref{ulas}).

{\it Proposition 4.4} The second Hamiltonian structure leads to the following
Poisson algebra
\ai
&&\{a_i, a_j\}_2={\rm (\ref{aiaj2})},\qquad i,j=1,2,\ldots,N-1;\\
&&\{a_{N}, a_{N}\}_2=\{u_0, u_0\}_2;\\
&&\{a_i, a_{N}\}_2=\{a_i, u_0\}_2,\quad 1\leq i\leq N-1;\\
&&\{S_1, S_1\}_2=\frac{N+1}{N}\ddxy.\label{s1s1}\\
&&\{a_j, S_1\}_2=\bigg[\frac{j}{N}\left(\ba{c}N+1\\ j+1\ea\right)
 \d^{j+1}+\sum_{l=1}^{j-1}\frac{(N+1)j-N(l+1)}{N(N-l)}
\left(\ba{c}N-l\\ N-j\ea\right) a_l\d^{j-l}\0\\\noal
&&\quad+\sum_{l=0}^{j-1}\d^{j-l}\bigg(
 \left(\ba{c}N\\ j-l-1\ea\right)\al_lS_1-
 \left(\ba{c}N+1\\ j-l\ea\right)\al_l'\bigg)
 +\sum_{l=0}^{j-1}(-1)^{j-l}\left(\ba{c} j\\ l\ea\right)
  \al_lS_1^{(j-l)}\label{ajs12'}\\\noal
&&\quad+\sum_{l=1}^{j-2}\sum_{k=0}^{j-l-2}a_l\d^{j-l-k-1}
  \bigg(\left(\ba{c}N-l\\ j-l-k-1\ea\right)\al_k'-
  \left(\ba{c}N-l-1\\ j-l-k-2\ea\right)\al_kS_1\bigg)\0\\\noal
&&\quad+\sum_{l=1}^{j-2}\sum_{k=0}^{j-l-2}(-1)^{j-l-k}
  \left(\ba{c}j-l-1\\ k\ea\right)a_l\al_kS_1^{(j-l-k-1)}\bigg]\dxy,\quad 1\leq
  j\leq N-1;\0\\\noal
&&\{a_N, S_1\}_2=\Bigl((\d+S_1)^N+\sum_{l=1}^{N-1}a_l(\d+S_1)^{N-l-1}
\Bigl)\ddxy.
\bj

The proofs of the above two propositions are not very difficult, so we skip
them.

\subsection{The general $W(N, M)$--algebra}

The Poisson algebras defined by the first Hamiltonian structure
is relatively simple. On the other hand the second Hamiltonian structure
certainly plays a more important role (see the Hamiltonian reduction below).
Hereafter we only pay attention to the second
Hamiltonian structure, and as we mentioned before,
we denote by $W(N, M)$ the finite algebra represented by the second Poisson
brackets of the fundamental fields of the $(N,M)$ model which is
encoded in (\ref{aiaj2}--\ref{uiuj2}). Contrary to the latter the $W(N,M)$
algebras are gauge--dependent, i.e. they depend on the coordinates
we choose for the model.
In Appendix {\bf B} we give
several explicit examples of the simplest $W(N, M)$ algebras.
In principle we can calculate any $W(N, M)$ algebra. Unfortunately
we cannot exhibit a compact explicit form of $W(N, M)$ with
arbitrary $N$ and $M$.
However it is not difficult to extract some general properties
of these algebras. This is the aim of the present subsection.

\vskip0.2cm
\noindent
\underline{\bf $(i)$. $W(N,M)$ in the gauge (\ref{pdo})}:
\vskip0.2cm

\noindent
In this gauge
\begin{itemize}
\item
$W(N, M)$ algebras are {\it local}\, and {\it polynomial}\,,
i.e. the Poisson brackets contain neither integration operators $\d^{-1}$ nor
fractional or negative powers of
the coordinates.
\item
$W(N, M)$--algebras are linear or bilinear in the $a_l$
fields but, in general, highly non--linear in the $S_l$ fields.
\item
The fields are characterized by their conformal spin; there are $(N+M-1)$
fields with spin ranging from $2$ to $(N+M)$;
in addition, there are $M$ spin one fields.
\item
For any Poisson bracket, the $\d^0$--terms in its RHS are either absent or
contain derivatives of the fields. In other words,
for any two coordinates $f$ and $g$, if
\a
\{f, g\}_2={\bf{\hat B}}\dxy,\qquad {\bf{\hat B}}=\sum_{l\geq0}b_l\d^l\0
\b
then either $b_0=0$ or $b_0$ contains derivatives of the
fundamental coordinates.
\end{itemize}
The first and the third assertions can be checked case by case, and in fact
they are true for all our examples in Appendix {\bf B} and
the previous subsection.
The second property can be obtained from the transformation (\ref{ulas})
and the algebra (\ref{aiaj2}--\ref{uiuj2}). Since
the algebra (\ref{aiaj2}--\ref{uiuj2}) is at most bilinear in the gauge
invariant functions, while the $u_l$ fields are linear in the $a_l$ fields and
non--linear in the $S_l$ fields, then $W(N, M)$ must be at most
bilinear in the $a_l$ fields but, in general highly non--linear in
the $S_l$ fields.

\vskip0.2cm
\noindent
\underline{\bf $(ii)$. $W(N,M)$ in the gauge (\ref{pdoqr})}:
\vskip0.2cm

\begin{itemize}
\item
$W(N, M)$ algebras are, in general,  {\it non-local}\, and
{\it non--polynomial}.

\item
Spin content: there are $(N-1)$ fields with integer spin running from
$2$ to $N$, and $2M$ fields with spins taking value
\a
\left\{\ba{l}
\relax[q_l]_{\rm conf}=\frac{N+1}{2},\\
\relax[r_l]_{\rm conf}=l+\frac{N-1}{2},
\ea\right.\qquad 1\leq l\leq M.\0
\b
\end{itemize}
In Appendix {\bf C} we will give some explicit examples of algebras in
this gauge.

\vskip0.2cm
\noindent
\underline{\bf $(iii)$. $W(N, M)$ algebras and $W_N$--algebras}
\vskip0.2cm

$W(N, M)$--algebras are intimately related to $W_N$--algebras in the
following sense:
\begin{itemize}
\item
They share the same Virasoro sub--algebra (\ref{vir}).
\item
The fields in the $W_N$--algebras are gauge invariant, while the
$W(N, M)$ algebras
contain all these gauge invariant fields plus some gauge dependent fields,
whose spins are integers or half--integers (depending on the gauge choice).
\item
(\ref{aiaj2}) is exactly a $W_N$--algebra modulo $u_l$--dependent terms.
\item
In the next section we will show that the $W(N, M)$ algebra can be reduced to
the usual $W_{N+M}$ algebra.
\end{itemize}

\vskip0.2cm
\noindent
\underline{\bf $(iv)$. Recurrences ~among~ $W(N, M)$ algebras}
\vskip0.2cm

{}From Appendix {\bf B},
we find that $W(1, 2)$ is almost a sub--algebra of $W(1, 3)$,
the only discrepancy being that $\{a_2, a_2\}_2$ in the latter case
contains  $a_3$--linearly--dependent terms.
This property holds for
two more pairs: $W(2,2)$ and $W(2, 1)$, $W(3,1)$ and $W_3$.
In fact, in general, we can write
\a
W_N {\subset}W(N,1){\subset}
\ldots {\subset}W(N, {\tilde M})
{\subset}W(N, M).\0
\b
This is to be interpreted in the following way:
for any ${\tilde M}<M$,  let us pick out from $W(N, {\tilde M})$
the Poisson brackets among $(a_1,\ldots,a_{N+{\tilde M}-1};
S_1,S_2,\ldots,S_{{\tilde M}})$, and mode out
$(a_{N+{\tilde M}},\ldots,a_{N+M-1};$\par\noindent
$S_{{\tilde M}+1},S_{{\tilde M}+2},\ldots,S_{M})$--dependent terms;
then we get the $W(N, {\tilde M})$--algebra.

\vskip0.2cm
\noindent
\underline{\bf $(v)$. $M$ dimensional subalgebras}
\vskip0.2cm

Beside the Virasoro algebra (\ref{vir}), $W(N, M)$ has another very simple
subalgebra, the Poisson algebra formed by the $S_i$ fields
\a
\{S_i, S_j\}_2=(\delta_{ij}+{1\over{N}})\ddxy.\label{sisjg}
\b
To derive it we notice that on a simple basis of dimensional counting the only
possible brackets are
\a
\{S_i, S_j\}_2={\rm const.}\ddxy,\label{possible}
\b
Then such form must remain unchanged when we take the dispersionless
limit (see below). In section {\bf 7}, when we analyse the dispersionless
$(N, M)$--th
\kdvh, we will prove that the constant is just $(\delta_{ij}+\frac{1}{N})$.
{}From this (\ref{sisjg}) follows in general.

Concerning this subalgebra let us make a side remark.
Let us define a new set of fields
\def\ts{\tilde S}
\a
\ts_1=S_1, \qquad \ts_j=S_j-S_{j-1},\qquad 2\leq j\leq M.\0
\b
Then the Poisson brackets among these new fields can be expressed in the
following matrix form
\a
\left(\ba{cccccccc} \frac{N+1}{N} &-1&0&\ldots&0&0&0\\
-1&2&-1&\ldots&0&0&0\\
\ldots&\ldots&\ldots&\ldots&\ldots&\ldots&\ldots\\
0&0&0&\ldots&-1&2&-1\\
0&0&0&\ldots&0&-1&2
\ea\right).
\b
This is almost the Cartan matrix of the $sl(M+1)$ Lie algebra,
except for the element in the upper left corner,
which is the Poisson bracket $\{\ts_1, \ts_1\}$. This may
hide some still unknown relation between $W(N, M)$ algebras and ordinary
finite dimensional Lie algebras.

\vskip0.2cm
\noindent
\underline{\bf $(vi)$. Some useful Poisson brackets}
\vskip0.2cm

{\it Proposition 4.5} In the $W(N, M)$ algebras, we find
in particular the following brackets:
\ai
&&\{a_j, S_1\}_2=\bigg[\frac{j}{N}\left(\ba{c}N+1\\ j+1\ea\right)
 \d^{j+1}+\sum_{l=1}^{j-1}\frac{(N+1)j-N(l+1)}{N(N-l)}
\left(\ba{c}N-l\\ N-j\ea\right) a_l\d^{j-l}\0\\\noal
&&+S_1-{\rm dependent~terms}\quad  j\leq N-1;\label{ajs12}\\\noal
&&\{a_N, S_1\}_2=\Bigl((\d+S_1)^N+\sum_{l=1}^{N-1}a_l(\d+S_1)^{N-l-1}
\Bigl)\ddxy;\label{ans12}\\\noal
&&\{a_{N+i}, S_j\}_2=0,\qquad i\geq j;\label{gaisj2}\\
&&\{S_i, S_j\}_2=\Bigl(\delta_{ij}+\frac{1}{N}\Bigl)\ddxy.\label{gsisj2}
\bj
These brackets are particularly important
since they are crucial for Hamiltonian reductions.

{\it Proof} : Let us start from eq.(\ref{gaisj2}): we see that it is true
for all the algebras given in Appendix {\bf B}.
Now let us consider the case of arbitrary $N$ and $M=2$ in which the gauge
invariant functions have the following realization
\def\tu{\tilde u}
\a
\tu_l=a_N(-\d+S_1)^l\cdot1
     +\sum_{l_1+l_2=l-1}a_{N+1}(-\d+S_2)^{l_2}(-\d+S_1)^{l_1}\cdot1.\0
\b
Comparing with eq.(\ref{ulas}), we see that
\a
\tu_l=u_l+\delta u_l, \qquad \delta u_l=a_{N+1}-{\rm dependent~terms}\0
\b
Since $\tu_l$'s and $u_l$'s (together with $\{a_l(1\leq l\leq N-1)\}$) form
the same algebra (\ref{aiaj2}--\ref{uiuj2}), we may use this invariance
to derive the additional Poisson brackets. For example
\a
\{\tu_0, \tu_1\}_2=\{u_0, u_1+a_{N+1}\}_2\0
\b
The $a_{N+1}$--dependent terms on the LHS determine $\{u_0, a_{N+1}\}_2
=\{a_N, a_{N+1}\}_2$. Similarly
\a
\{\tu_1, \tu_1\}_2-\{u_1, u_1\}_2=\{a_{N+1}, u_1\}_2
+\{u_1, a_{N+1}\}_2+\{a_{N+1}, a_{N+1}\}_2,\0
\b
The RHS must be $a_{N+1}$--dependent, since all the $a_{N+1}$--independent
terms
are excluded. This requirement implies that
\a
\{a_{N+1}, S_1\}_2\quad {\rm must~be}\quad a_{N+1}-{\rm dependent}\0
\b
However, by dimension counting, it is easy to see that $a_{N+1}$--dependent
terms are not allowed in local polynomial Poisson brackets.
Therefore we must have
\a
\{a_{N+1}, S_1\}_2=0.\0
\b

For the same reason, when we consider $M\geq 3$, i.e. we add more
pseudo--differential terms to the Lax operator, we will obtain
(\ref{gsisj2}).
Therefore, when we add more \pdo\, terms
to the Lax operator, they will not change the Poisson brackets involving
the $S_l$ fields already considered, thus we must have in general
(\ref{ajs12}).

\vskip0.2cm
\noindent
\underline{\bf $(vii)$. The generating algebras of the usual $W$--infinity
algebras}
\vskip0.2cm

Finally let us consider the $N=1$ case. As we have pointed out several times
the gauge invariant functions $\{u_l\}$ form a $W_{1+\infty}$ and
a $W_\infty$ algebra. However these two algebras can be realized also
by means of $2M$ fields only, via the combinations (\ref{ulas}) and the
finite dimensional algebras determined by the first and second Poisson
brackets of the fields. We could therefore phrase this situation by saying
that the $W_{1+\infty}$ and the $W_{\infty}$ algebras reduce to
the latter finite algebras.

Since $M$ can be any positive
integer we see that there are infinite many different realizations of the
$W_{1+\infty}$
and the $W_\infty$ algebras. For instance, $W(1,2)$ and $W(1,3)$ give
the four-- and six--field realizations of the $W_\infty$ algebra.
Alternatively we can say that $W_{1+\infty}$ and
$W_\infty$ algebras are highly reducible, and multi--field representations
of the \kph\, provide a way to classify the reduced algebras.

\section{Reductions of the $(N, M)$--th \kdvh}

\setcounter{equation}{0}
\setcounter{subsection}{0}
\setcounter{footnote}{0}

We have claimed above that $W(N, M)$ algebra can be reduced to the usual
$W_{N+M}$ algebra. In this section we will show this and the more general
relations (\ref{kdvred}) and (\ref{algred}). Let us first summarize
the two schemes for Hamiltonian reduction with second class constraints
we will be using.

\subsection{Two reduction schemes for Hamiltonian systems}

Let us consider a Hamiltonian system, no matter whether it is integrable
or non--integrable. It will have fields generically denoted by $f_i$, a
Hamiltonian $H$ and a Poisson bracket.
The equations of motion are
\a
\frac{\d}{\d t} f_i=\{f_i, H\}, \qquad i=1,2,\ldots,n.\0
\b
Now we want to impose, for example, $C=0$
where $C$ is a particular combination of the fields and their derivatives,
and suppose that it is
second class, i.e.
\a
\{C, C\}=\Delta\dxy, \qquad \Delta|_{C=0}\neq0,\0
\b
then, in order to avoid inconsistencies, we can proceed in two ways.

\vskip3mm
\noindent
{\bf The first scheme}
\vskip2mm

If we explicitly know all the Poisson brackets among the independent fields,
we can first improve the Poisson bracket, then derive the reduced
equations of motion.

{\it step 1} :Introduce in the reduced phase space the Dirac--Poisson bracket
\a
\{f_i, f_j\}_{\rm D}\stackrel{\rm def}{=}\Bigl(
\{f_i, f_j\}-\{f_i, C\}\frac{1}{\Delta}\{C, f_j\}\Bigl)_{C=0}.
\label{dirac}
\b

{\it step 2} :The Hamiltonian of the reduced system is just
$H|_{f_1=0}$. Using this restricted Hamiltonian and the Dirac--Poisson bracket,
we are able to derive the equations of motion for the reduced Hamiltonian
system.

\vskip3mm
\noindent
{\bf The second scheme}
\vskip2mm

If it so happens that we do not know the full Poisson algebra, but we know the
Poisson brackets between the constraint and all the fields, then we can
improve the Hamiltonian with the addition of a suitable Lagrange
multiplier term, and make use of the known Poisson brackets to derive
the equations of motion of the reduced system.

{\it step 1} : Add a Lagrange multiplier term to $H$
\def\th{\tilde H}
\a
H\Longrightarrow \th=H+\int \alpha f_1,\0
\b
where $\al$ is an expression to be determined.

{\it step 2} : Using this new Hamiltonian and the original Poisson brackets
derive the equation of motion for $C$,
\a
\frac{\d}{\d t}C=\{C, \th\}.\0
\b

{\it step 3} : Determine $\al$ so that
the second class constraint is preserved by the time evolution generated
by the new Hamiltonian, i.e. turn $C=0$ into a first class constraint.

{\it step 4} : Using this new Hamiltonian and the original Poisson brackets
derive the equations of motion for the reduced system
\a
\frac{\d}{\d t}f_i=\{f_i, \th\}|_{C=0},\qquad i=2,3,\ldots,n.\0
\b

\vskip3mm
\noindent
{\bf Comment concerning integrable Hamiltonian systems}
\vskip2mm

What we said so far is valid for any Hamiltonian system.
Now let us turn our attention to the reduction of an integrable Hamiltonian
system. After implementing either of the above two schemes, we have to make
sure that the reduced system is still integrable. In other words we should
further
construct its bi--Hamiltonian structure, which is a recursive but lengthy
procedure. A shortcut in this sense consists in
finding its Lax pair representation.

Let us finally remark that the two schemes presented above lead to the
same results, to the extent we have been able to produce explicit results
for both of them.

\subsection{Reduction of $W(N,M)$ algebras}

We want to impose the constraint $S=0$ for some of the $S$ fields of the
algebra, and this is second class.

Applying the first reduction scheme to the $W(1,2)$--algebra, one can prove
the following sequence of reductions \cite{bx3}
\a
W(1,2)\stackrel{S_1=0}{{\longrightarrow\longrightarrow} }W(2, 1)
\stackrel{S_2=0}{\longrightarrow\longrightarrow}W(3, 0)=W_3.\0
\b
For the algebras given in Appendix {\bf B}, we can find another
sequence
\a
W(1, 3)\stackrel{S_1=0}{\longrightarrow\longrightarrow}W(2, 2)
\stackrel{S_2=0}{\longrightarrow\longrightarrow}W(3, 1)
\stackrel{S_3=0}{\longrightarrow\longrightarrow}W(4, 0)=W_4.\0
\b
These two sequences confirm the relation (\ref{algred}).
However, since we do not have at hand a compact explicit form of
the algebra $W(N,M)$ we can verify that with $S_1=0$, $W(N,M)$ reduces
to $W(N+1,M-1)$ only case by case. To overcome this difficulty we have
to resort to the second reduction scheme.

\subsection{Reduction of the hierarchy}

{\it Proposition 5.1} : When we impose the constraint $S_1=0$,
the $W(N, M)$--algebra
reduces to the $W(N+1, M-1)$--algebra
\a
W(N,M) \stackrel{S_1=0}{\longrightarrow\longrightarrow}W(N+1,
M-1)\label{algred'}
\b
and, simultaneously, the $(N,M)$ hierarchy
reduces to the $(N+1, M-1)$ hierarchy.

{\it Proof} : As anticipated above, we adopt the second reduction scheme.
Our starting point is
eqs.(\ref{gkdv}--\ref{ppkdv}), which are generated by the second Hamiltonian
$H_2=\frac{N}{2}<L^{\frac{2}{N}}>$ through the second Poisson structure.
Obviously the constraint $S_1=0$ is not preserved by the time evolutions, so
we modify $H_2$ as follows
\def\th{\tilde H}
\a
H_2\Longrightarrow \th_2=H_2+\int \alpha S_1,\0
\b
with $\al$ to be determined. This new Hamiltonian generates
the following equation of motion for the $S_1$ field (with respect to the
original second Poisson bracket)
\a
\Bigl(\ddt 2 S_1\Bigl)_{\rm improved}=
\frac{2}{N}a_1'+2S_1S_1'-S_1^{''}+\frac{N+1}{N}\al+\int\{S_1, \al\}_2S_1.\0
\b
Setting $S_1=0$, we obtain
\a
\al=-\frac{2}{N+1}a_1\0
\b
therefore our improved Hamiltonian is
\a
\th_2=2\int\bigg(a_2(x)+
\Big(\delta_{N,1}-\frac{1}{N+1}\Big)a_1(x)S_1(x)\bigg) dx.
\label{imph2}
\b
Using eq.(\ref{ajs12}) and
(\ref{gsisj2}), we can derive the explicit form of the equations of motion
of the reduced system
\ai
&&\ddt 2 a_l=a_l^{''}+2a'_{l+1}-{2\over{N+1}}
  \left(\ba{c}N+1\\l+1\ea\right)a_1^{(l+1)}\0\\\noal
&&\qquad\qquad-{2\over{N+1}}\sum_{k=1}^{l-1}
  \left(\ba{c}N-k\\l-k\ea\right)a_ka_1^{(l-k)},
   \qquad 1\leq l\leq N;\label{gkdv'}\\\noal
&&\ddt 2 a_{N+l}=a^{''}_{N+l}+2a'_{N+l+1}
  +2a'_{N+l}S_{l+1}+2a_{N+l}\Bigl(\sum_{k=1}^{l+1} S_k\Bigl)',
  ~~1\leq l\leq M-1;\label{pkdv'}\\
&&\ddt 2 S_l={2\over{N+1}}a_1'+2S_lS_l'-S_l^{''}
  -2\Bigl(\sum_{k=1}^{l-1} S_k\Bigl)^{''},\qquad   2\leq l\leq M.
\label{ppkdv'}
\bj
We immediately recognize that this is nothing but the first
non--trivial flow of the $(N+1, M-1)$--th \kdvh.
The only effect of our reduction is the shift
\a
N{\longrightarrow\longrightarrow}N+1,
\qquad M{\longrightarrow\longrightarrow}M-1.\0
\b
The algebra associated to this hierarchy is $W(N+1, M-1)$, thus we are led
to the relation (\ref{algred'}).

But now we remark that our proof is valid for any value of $M$. Therefore
we can adapt the above proof to the case
when we set $S_2=0,
S_3=0,\ldots, S_M=0$. Finally we get the $(N+M)$--th \kdvh\, and the $W_{N+M}$
algebra.

\section{Drinfeld--Sokholov representation}

\setcounter{equation}{0}
\setcounter{subsection}{0}
\setcounter{footnote}{0}

The integrable models and reductions we have been considering so far
can be synthesized in a very compact and useful form via suitable
Drinfeld--Sokholov (DS) linear systems. From
them one can easily extract the corresponding Lax pair.

\subsection{The linear system associated to the $(N, M)$--th \kdvh}

The $(N, M)$--th \kdvh\, (\ref{gkdvh}) can be viewed as the consistency
condition of the following linear system
\a
\left\{\ba{l}
L\Psi_0(\lm,t)=\lm \Psi_0(\lm, t),\\\noal
\ddt r \Psi_0(\lm,t)=\Bigl(L^{r\over{N}}\Bigl)_+\Psi_0(\lm,t)
\ea\right.\label{psi}
\b
where $\lm$ is the spectral parameter, and $\Psi_0$ is referred to as
Baker--Akhiezer function. In \cite{bx2} we have shown that this linear
system naturally appears in multi--matrix models.

In terms of Baker--Akhiezer function, we can introduce an important
ingredient -- \tauf,
\a
\Psi_0(\lm,t)=\frac{V(\lm, t)\tau(t)}{\tau(t)},\label{tau}
\b
where
\a
V(\lm, t)=\exp{(\sum_{r=1}^\infty t_r\lm^r)}
\exp{(-\sum_{r=1}^\infty{1\over{r\lm^r}}{\d\over{\d t_r}})}\0
\b
is a vertex operator. One can show that the \tauf\, satisfies
the following relations \cite{dickey}
\a
{{\d^2}\over{\d t_1\d t_r}}\ln \tau(t)=\res_{\d} L^{r\over{N}}.\label{taul}
\b

\subsection{The generalized DS representation}

Now let us return to the spectral problem (\ref{psi}).
The first equation of (\ref{psi}) is highly non--linear,
our aim is to linearize it. In order to do so,
we introduce some more notations,
\a
(E_{ij})_{kl}=\delta_{ik}\delta_{jl}, \qquad 1\leq i,j,k,l\leq N+M,\0
\b
which is the $(N+M)\times(N+M)$ matrix with only one non--zero element
at the position $(i,j)$. Define
\a
&&I_+\equiv \sum_{i=1}^{N+M-1}E_{i,i+1},\0\\
&&{\bf \Psi}\equiv
(\Psi_{-M},\Psi_{1-M},\ldots,\Psi_{-1},\Psi_0,\Psi_{1},\ldots,
\Psi_{N})^T,\\
&& {\bf S}\equiv \sum_{i=1}^M S_{M-i+1}E_{i,i},\qquad\quad
 {\bf A}\equiv \sum_{i=1}^{N+M-1} a_{i}E_{N+M,N+M-i}.\0
\b
So ${\bf \Psi}\equiv {\bf \Psi}(\lambda , t)$ is a column vector, the other are
matrices. ${\Psi_0}$ is the Baker--Akhiezer function of eq.(\ref{psi}). Now
we can express the spectral equation in linear form as follows
\a
\Bigl(\d+{\bf S}+{\bf A}
-\lm E_{N+M,M+1}-I_+\Bigl){\bf \Psi}=0.\label{gds}
\b
If we eliminate all the $\Psi_l$ in favor of $\Psi_0$, we recover
the spectral equation in (\ref{psi}).

Now let us consider some particular cases

\noindent
$(i)$. $M=0$ case, this is the $N$--th \kdvh, the above linear
system is just the original Drinfeld--Sokholov representation, in which the
spectral parameter $\lm$ is put on the lower left corner \cite{ds}.

\noindent
$(ii)$. $N=1$ case, this is $2M$--field representation of \kph. The spectral
parameter in this case is located on the main diagonal line (on the lower
right corner).

For general `$N$' and `$M$', the spectral parameter appears in the
last row. When we make the reductions discussed in section {\bf 5}, Imposing
$S_1=0$ is equivalent to replacing
\a
N\longrightarrow N+1, \qquad M\longrightarrow M-1.\0
\b
In (\ref{gds}), this change corresponds to moving spectral
parameter $\lm$ from the right to the left by one step. Repeating the
procedure,
we finally can move $\lm$ to the lower left corner, that is, we get a
\kdvh.

\section{The dispersionless $(N, M)$--th KdV hierarchy}

\setcounter{equation}{0}
\setcounter{subsection}{0}
\setcounter{footnote}{0}

In this section we will consider the dispersionless limit of the
$(N, M)$--th \kdvh\,
(\ref{gkdvh}). The usual procedure to get this limit is the following:
we simply ignore the higher than first
derivatives in the equations of motion as well as in the Poisson brackets.
This is equivalent to substituting the commutators by the basic
canonical Poisson relation, i.e.
\a
[\d, x]\Longrightarrow \{p, x\}=1,\label{px}
\b
where $p$ denotes the canonical conjugate momentum. In terms of this elementary
Poisson structure, the dispersionless $(N, M)$--th KdV hierarchy can be
written as \def\l{{\cal L}}
\a
\ddt r \l=\{\l^r_+, \l\}, \label{disgkdvh}
\b
with the dispersionless Lax operator
\a
\l&=&p^{N}+\sum_{l=1}^N a_l p^{N-l}+\sum_{l=1}^M
   a_{N+l-1}{1\over{(p-S_1)(p-S_2)\ldots(p-S_l)}}\0\\
&=&p^{N}+\sum_{l=1}^N a_l p^{N-l}+\sum_{l=1}^\infty u_l p^{-l-1}.
\label{dislax}
\b
The subscript `$+$' in (\ref{disgkdvh}) means that we keep only non--negative
powers of `$p$'.
As an example we give the second dispersionless flow equations
\ai
&&\ddt 2 a_l=2a'_{l+1}-{{2(N-l)}\over{N}}
a_{l-1}a_1',   \qquad 1\leq l\leq N-1;\label{disgkdv}\\
&&\ddt 2 a_{N+l}=2a'_{N+l+1}
  +2a'_{N+l}S_{l+1}+2a_{N+l}\Bigl(\sum_{k=1}^{l+1} S_k\Bigl)', \quad
  0\leq l\leq M-1;\label{dispkdv}\\
&&\ddt 2 S_l={2\over{N}}a_1'+2S_lS_l',\qquad   1\leq l\leq M.
\label{disppkdv}
\bj
Comparing with eqs(\ref{gkdv}--\ref{ppkdv}), we see that only the first
order derivatives survive.
In the dispersionless limit, the gauge invariant functions $u_l$'s
have very simple expressions. The first few are
\a
&&u_0=a_{N},\qquad\qquad u_1=a_{N+2}+a_{N}S_1,\0\\
&&u_2=a_{N+3}+a_{N+2}(S_1+S_2)+a_{N}S_1^3,\0
\b
In order to find general compact formulas we introduce
a set of {\it completely symmetric} and {\it homogeneous} polynomials $e_k$
\a
e_k(S_1, S_2,\ldots, S_M)\stackrel{\rm def}{=}
   \sum_{1\leq i_1\leq i_2\leq\ldots\leq M}
  S_{i_1}S_{i_2}\ldots S_{i_k},\qquad k\in{\bf Z}_+\label{pk}
\b
In particular we have
\a
e_k(S_1, S_2,\ldots, S_M)=
  \sum_{l=0}^k S_1^le_{k-l}(S_2, S_3,\ldots, S_M).\label{pks1}
\b
Without further specification hereafter we understand
$e_k=e_k(S_1,S_2,\ldots,S_M)$. It is not difficult to show that
these symmetric functions satisfy the following identities(see Appendix
{\bf D} for the proofs)
\ai
&&\frac{\d}{\d S_l}e_{i+1}=\sum_{\mu=0}^iS_l^{\mu}e_{i-\mu};\label{dsp}\\
&&\sum_{l=1}^M\sum_{\mu=0}^iS_l^{\mu}e_{i-\mu}=(i+M)e_i;\label{sumsp}\\
&&e'_{i+1}\stackrel{\rm def}{=}\sum_{l=1}^MS'_l\frac{\d}{\d S_l}e_{i+1}
=\sum_{l=1}^M\sum_{\mu=0}^iS'_lS_l^{\mu}e_{i-\mu};\label{p'}\\
&&\sum_{l=1}^M\sum_{\beta=0}^{i-1}\sum_{\alpha=\beta+1}^i
 S^\beta_l e_{i-\alpha}e_{j+\alpha-\beta}=\sum_{\alpha=1}^i(i-\alpha+M)
 e_{i-\alpha}e_{j+\alpha};\label{pij-+}\\
&&\sum_{l=1}^M\sum_{\alpha=1}^i\sum_{\beta=0}^{\alpha-1}
 (\alpha-\beta-1)S_l'S^\beta_l e_{i-\alpha}e_{j+\alpha-\beta-1}=
 \sum_{\alpha=1}^i\alpha e'_{i-\alpha}e_{j+\alpha};\label{pi'j-+}\\
&&\sum_{l=1}^M\Bigl(\frac{\d}{\d S_l}e_{i+1}\Bigl)\d\Bigl(
 \frac{\d}{\d S_l}e_{j+1}\Bigl)
=(j+M)e_{i}\d e_{j}
+\sum_{l=1}^i\Bigl(le_{j+l}\d e_{i-l}
+(j-i+l)e_{i-l}\d e_{j+l}\Bigl).\label{dspdsp}
\bj
The last one is an operatorial equation.
Using these symmetric polynomials we obtain
\a
&&u_l=\sum_{k=1}^{l+1}a_{N+k-1}e_{l-k+1}(S_1, S_2, \ldots, S_k),
  \qquad 0\leq l\leq M-1;\0\\
&&u_l=\sum_{k=1}^M a_{N+k-1}e_{l-k+1}(S_1, S_2, \ldots, S_k),
  \qquad l\geq M.\label{ulpk}
\b
In the dispersionless limit the two Hamiltonian structures are extremely
simplified. With a little exercise, we get the first Poisson algebra
\ai
&&\{a_i, a_j\}_1=\Bigl(N\delta_{i+j,N}\d+(N-i)a_{i+j-N-1}\d+
(N-j)\d a_{i+j-N-1}\Bigl)\dxy,\\
&&\{u_i, u_j\}_1=(iu_{i+j-1}\d+j\d u_{i+j-1})\dxy,\\
&&\{a_i, u_j\}_1=0.
\bj
The second Poisson algebra is
\ai
&&\{a_i, a_j\}_2=\big[ia_{i+j-1}\d+j\d a_{i+j-1}+iu_{i+j-N-1}\d+
   j\d u_{i+j-N-1}\0\\
&&\quad +\sum_{l=1}^{i-1}\Bigl((i-l-1)a_{i+j-l-2}\d a_l
        +(j-l-1)a_l\d a_{i+j-l-2}\Bigl)\label{disaiaj2}\\
&&\quad +\sum_{l=1}^{i-2}\Bigl((j-l-1)a_l\d u_{i+j-l-N-2}
        +(i-l-1)u_{i+j-l-N-2}\d a_l\Bigl)\0\\
&&\quad+\frac{(N-i)(N-j)}{N} a_{i-1}\d a_{j-1}\big]\dxy,\0\\
&&\{u_i, u_j\}_2=\big[(N+i)u_{i+j+N-1}\d+(N+j)\d u_{i+j+N-1}
+\frac{ij}{N} u_{i-1}\d u_{j-1}\0\\
&&\quad +\sum_{l=1}^{N-1}\Bigl((N+j-l-1)a_l\d u_{i+j+N-l-2}
        +(N+i-l-1) u_{i+j+N-l-2}\d a_l\Bigl)\label{disuiuj2}\\
&&\quad +\sum_{l=0}^{i-1}\Bigl((i-l-1) u_{i+j+N-l-2}\d u_l+
        +(j-l-1)u_l\d  u_{i+j+N-l-2}\Bigl)\big]\dxy,\0\\
&&\{a_i, u_j\}_2=\big[iu_{i+j-1}\d+(N+j)\d u_{i+j-1}
   -\frac{(N-i)j}{N} a_{i-1}\d u_{j-1}\label{disaiuj2}\\
&&\quad +\sum_{l=1}^{i-2}\Bigl((i-l-1)u_{i+j-l-2}\d a_l
        +(N+j-l-1)a_l\d u_{i+j-l-2}\Bigl)\big]\dxy.\0
\bj
The conserved quantities are
\footnote{Here the residue means simply
the coefficient of the $p^{-1}$ term.}
\a
H_r=\frac{N}{r}\int \res_{p}\Bigl(\l^{r\over N}\Bigl),\qquad r\geq1.
\b
One can check that the two Poisson brackets are compatible w.r.t.
these quantities, and they indeed generate the flows (\ref{disgkdvh}).
In the remaining part of this section we will try to derive the Poisson
brackets among different $S_l$'s, and the Poisson relations between $S_1$
and other fields.

{\it Proposition 7.1} The symmetric polynomials $e_i$ satisfy
the following Poisson brackets
\a
\{e_{i+1}, e_{j+1}\}_2&=&\Bigl[\Bigl(\frac{i+M}{N}+1)(j+M)e_i\d e_j
\label{pipj2}\\
&&+\sum_{l=1}^i\Bigl(le_{j+l}\d e_{i-l}
+(j-i+l)e_{i-l}\d e_{j+l}\Bigl)\Bigl]\dxy\0
\b

{\it Proof} :
As we discussed in subsection {\bf 4.3} the possible Poisson brackets
among the $S_l$ fields are of the form (\ref{possible}). Taking the Poisson
bracket between two monomials of the
$S_l$ fields reduces the total number of $S_l$
fields by $2$. Eq.(\ref{ulas}) shows that in $u_{i+M}(i\geq0)$
the terms with lowest powers of  $S_l$ fields are $a_{N+M-1}e_{i+1}$.
Now let us consider the Poisson bracket
$\{u_{i+M}, u_{j+M}\}_2$ with $i,j\geq0$. In this
expression the terms with lowest powers of $S_l$ fields will come from
\a
a_{N+M-1}\{e_{i+1}, e_{j+1}\}_2a_{N+M-1}\0
\b
On the other hand, the comparable terms in the RHS of the Poisson bracket
(\ref{disuiuj2}) are
\a
&&a_{N+M-1}\Bigl[\Bigl(\frac{i+M}{N}+1\Bigl)(j+M)e_i\d e_j
  +\sum_{l=1}^i\Bigl(l e_{j+l}\d e_{i-l}\0\\
&&+(j-i+l)e_{i-l}\d e_{j+l}\Bigl)\Bigl]a_{N+M-1}\dxy\0
\b
Comparing these two expressions, we obtain (\ref{pipj2}).

{\it Proposition 7.2}  The Poisson brackets (\ref{pipj2}) imply the Poisson
algebra (\ref{gsisj2}).

{\it Proof}. This is not difficult to prove. We have already noticed that the
bracket $\{S_1, S_j\}$ must have the form (\ref{possible}).
Now making use of the relations
(\ref{pks1}) and (\ref{dsp}--\ref{dspdsp}), we can prove that (\ref{gsisj2})
indeed leads to eqs.(\ref{pipj2}). This uniquely fixes the undetermined
constant
of (\ref{possible}).

The Poisson brackets between $S_1$ and $a_l(1\leq l\leq N+M-1)$
are easier to derive. In fact we can get them directly
from eq.(\ref{ajs12}--\ref{gsisj2}) by suppressing all higher order
derivatives, the result is
\ai
&&\{a_i, S_1\}_2=\big[S_1^i+\sum_{l=1}^{i-2}a_lS_1^{i-l-1}+\frac{i}{N}
  a_{i-1}\big]\ddxy,\quad 1\leq i\leq N-1;\label{disais12}\\
&&\{a_{N}, S_1\}_2=\Bigl(S_1^{N}+\sum_{l=1}^N a_lS_1^{N-l}\Bigl)\ddxy;
 \label{disan1s12}\\
&&\{a_{N+l-1}, S_1\}_2=0,\qquad l=2,3,\ldots, M.\label{disan1sl2}
\bj

\section{Reductions of the dispersionless $(N, M)$--th KdV hierarchy}

\setcounter{equation}{0}
\setcounter{subsection}{0}
\setcounter{footnote}{0}

Starting from the previous results we will examine in this section
the reduction of the
dispersionless $(N, M)$--th \kdvh\,. To find reductions we may suppress
the fields $S_l$ one by one. This was already done in section {\bf 5} and in
\cite{bx3} for the dispersive (general) case and will not be repeated here.
In this section we are interested in the possible existence of other
reductions.
Let us take the dispersionless four--boson representation of KP hierarchy as
an example. The Lax operator is
\a
\l_1=p+\frac{a_1}{p-S_1}+\frac{a_2}{(p-S_1)(p-S_2)}.
\label{l02'}
\b
The second Poisson algebra can be obtained from (\ref{w31}) by killing
all the higher derivatives
\a
&&\{a_1, a_1\}_2=(a_1\d+\d a_1)\delta(x-y),
\quad\quad\{a_1, a_2\}_2=(a_2\d+2\d a_2)\delta(x-y),\0\\
&&\{a_1, S_1\}_2=S_1\ddxy,
\quad\quad\{a_1, S_2\}_2=S_2\delta'(x-y),\0\\
&&\{a_2, a_2\}_2=[a_2(2S_2-S_1)\d+\d a_2(2S_2-S_1)]
\delta(x-y),\label{disw31}\\
&&\{a_2, S_2\}_2=\Bigl(a_1+S_2(S_2-S_1)\Bigl)
\delta'(x-y),\0\\
&&\{a_2, S_1\}_2=0,\quad
  \{S_i, S_j\}_2=(\delta_{ij}+1)\delta'(x-y),\quad i,j=1,2.\0
\b
We call it the classical $w(1,2)$ algebra.
With respect to this Poisson algebra, the second Hamiltonian
\a
H_2=\int (a_2+a_1S_1), \label{disl02h2}
\b
will generate the first non--trivial flow equations
\a
&&\ddt 2 a_1=2(a_2+a_1S_1)',\0\\
&&\ddt 2 a_2=2a'_2S_2+2a_2(S_1+S_2)',\0\\
&&\ddt 2 S_1=(2a_1+S_1^2)',\0\\
&&\ddt 2 S_2=(2a_1+S_2^2)'.\0
\b

Now, instead of imposing $S_1=0$ as we have done previously,
we let $S_1=S_2$.
Starting from the algebra (\ref{disw31}) we get an
improved algebra
\a
&&\{a_1, a_1\}=\Bigl(a_1\d+\d a_1\Bigl)\dxy,\qquad
\{a_1, a_2\}=(a_2\d+2\d a_2)\dxy,\0\\
&&\{a_2, a_2\}=\Bigl(a_2S\d+\d a_2S-\ha a_1\d a_1\Bigl)\dxy,\label{w32''}\\
&&\{a_1, S\}=S\ddxy,\quad
\{a_2, S\}=\ha a_1\ddxy,\quad
\{S, S\}={3\over2}\ddxy.\0
\b
With respect to this reduced algebra,
the original Hamiltonian (\ref{disl02h2}) generates
the following flow equations
\a
&&\ddt 2 a_1=2(a_2+a_1S)',\0\\
&&\ddt 2 a_2=2a'_2S+4a_2S',\0\\
&&\ddt 2 S_2=(2a_1+S^2)'.\0
\b
It turns out that these equations
admit the following Lax representation \footnote{This implies integrability for
the reduced system. The $S_1=S_2$ reduction for the corresponding dispersive
hierarchy was studied in \cite{bx3}; the reduced system however is not
integrable,
at least as long as we stick to locality and polynomiality.}
\a
\ddt r \l_4=\{(\l^r_4)_+, \l_4\}.\0
\b
with
\a
\l_2=p+\frac{a_1}{p-S}+\frac{a_2}{(p-S)^2}.\label{l4}
\b
As we explained in \cite{bx3}, imposing $S=0$ will reduce the above
hierarchy to the dispersionless Boussinesq hierarchy.

The obvious generalization of the $S_1=S_2$ reduction to other hierarchies
consists
in picking out $i~(i=1, ..., M)$ of the $S_l$ fields and setting them equal.
So altogether we expect to find $k$ distinct reductions with
\a
k=\left(\ba{c}M\\2\ea\right)+\left(\ba{c}M\\3\ea\right)+\ldots+
\left(\ba{c}M\\M\ea\right)=2^M-M-1.\0
\b
We have checked this for the $6$--boson
field representation of \kph.

\section{Conclusions}

\setcounter{equation}{0}
\setcounter{subsection}{0}
\setcounter{footnote}{0}

The content of our paper can be summarized as follows.
We have systematically discussed the $(N, M)$--th \kdvh\,.
The large integrable differential hierarchy (\ref{gkdvh}) contains
both the higher \kdvh\, and the multi--field representations of \kph.
The \shs\, of this hierarchy leads to the extended $W$--algebra,
$W(N,M)$. By suppressing the $S_l$ fields in succession  we can
reduce the $W(N, M)$ algebra to the usual $W_{N+M}$--algebra. Simultaneously
the corresponding $(N, M)$--th \kdvh\, reduces to $(N+M)$--th \kdvh.
There do not seem to exist other integrable reductions
as long as we put restrictions only on the fields $S_l$.
However in the dispersionless limit there do exist
another kind of reduction in which we identify some of the $S_l$
fields.  The resulting integrable hierarchy can be obtained by directly
imposing this constraint in the Lax operator.

In \cite{bxII} we use the results of this papers to compute
correlation functions of the two--matrix models. As far as the study
of hierarchies and algebras is concerned, there remain some open questions.
One is whether there are other reductions of the $(N,M)$ hierarchy of
different nature than the ones we have considered.
Another problem concerns the quantum versions of the algebras $W(N,M)$.
We have seen in section 4 that each algebra $W(N, M)$
contains a Virasoro sub--algebra (\ref{vir}); this may suggest the
quantum version of $W(N, M)$ algebra to correspond to some new
conformal field theory. \cite{lx} contains some hints on this direction:
the $W(2,1)$ algebra is examined and its several free field realizations and
quantum version are derived.
%The resulting quantum $W(1,2)$ algebra has negative conformal anomaly.

A third interesting question is whether our $W(N, M)$ algebra has any relations
with the $W_N^{(l)}$ introduced in \cite{bakas}.
The latter is deduced from a WZW model via Hamiltonian reduction.
Finally it would be quite interesting
to understand the relation between the extended \kdvh\, (\ref{gkdvh}) and
the conformal affine Toda theories (CAT models),  for we know
the two boson representation
of \kph naturally appears in the $sl(2)$ CAT model.

\vskip 1.5cm
\noindent
{\Large\bf Appendices}
\vskip1cm

\appendix

\section{Derivation of the second flow equations}

\setcounter{equation}{0}
\setcounter{subsection}{0}
\setcounter{footnote}{0}

In this Appendix we will derive the second flow equations
eqs.(\ref{gkdv}--\ref{ppkdv}). First we see that
\a
\ddt 2 L&=&\sum_{l=1}^{N-1}\Bigl({{\d a_l}\over{\d t_2}}\Bigl)\d^{N-l-1}
  +\sum_{l=1}^M\Bigl({{\d a_{N+l-1}}\over{\d t_2}}\Bigl)
{1\over{\d-S_l}}{1\over{\d-S_{l-1}}}\ldots{1\over{\d-S_1}}\0\\
&+&\sum_{l=1}^M\sum_{k=1}^l
a_{N+l-1}{1\over{\d-S_l}}\ldots{1\over{\d-S_k}}\Bigl({{\d S_k}\over{\d
t_2}}\Bigl)
{1\over{\d-S_k}}\ldots{1\over{\d-S_1}}.\label{2ndf}
\b
Our next task is to calculate the commutator
$[\Bigl(L^{2\over{N}}\Bigl)_+, L]$. Obviously, we have
\a
\Bigl(L^{2\over{N}}\Bigl)_+=\d^2+{2\over{N}}a_1,\0
\b
After a straightforward calculation we obtain
\a
&&[\Bigl(L^{2\over{N}}\Bigl)_+, L_+]
 =[\d^2+{2\over{N}}a_1, \d^{N}+\sum_{l=1}^{N-1} a_l\d^{N-l-1}]\0\\\noal
&&\qquad=2\sum_{l=1}^{N-1} a'_l\d^{N-l}+\sum_{l=1}^{N-1} a_l^{''}\d^{N-l-1}
  -{2\over{N}}\sum_{r=1}^{N}\left(\ba{c}N\\r\ea\right)
  a_1^{(r)}\d^{N-r}\0\\\noal
&&\qquad-{2\over{N}}\sum_{l=1}^{N-2}a_l\sum_{r=1}^{N-l-1}
  \left(\ba{c}N-l-1\\r\ea\right)a_1^{(r)}\d^{N-r-l-1},\0\\\noal
&&[\d^2+{2\over{N}}a_1, a_{N+l-1}]=2a'_{N+l-1}\d+a_{N+l-1}^{''}\0\\\noal
&&\qquad =2a'_{N+l-1}(\d-S_l)+2a'_{N+l-1}S_l+a_{N+l-1}^{''},\0\\\noal
&&[\d^2+{2\over{N}}a_1, \d-S_k]
 =-2S'_k\d-S_k^{''}-{2\over{N}}a_1'\label{c1}\\\noal
&&\qquad=-(\d-S_k)2S_k'+S_k^{''}-2S_kS'_k-{2\over{N}}a_1',\0\\\noal
&&[\d^2+{2\over{N}}a_1, {1\over{\d-S_k}}]
 =-{1\over{\d-S_k}}[\d^2+{2\over{N}}a_1, \d-S_k]
  {1\over{\d-S_k}}\0\\\noal
&&\qquad=2S_k'{1\over{\d-S_k}}+{1\over{\d-S_k}}
 \bigg({2\over{N}}a_1'+2S_kS'_k-S_k^{''}\bigg){1\over{\d-S_k}}.\0
\b
The last two formulas enable  us to calculate
\a
&&[\Bigl(L^{2\over{N}}\Bigl)_+,
 +\sum_{l=1}^M a_{N+l-1}{1\over{\d-S_l}}{1\over{\d-S_{l-1}}}
  \ldots {1\over{\d-S_1}}]\0\\\noal
&&=2a'_{N}+2\sum_{l=1}^{M-1} a'_{N+l-1}{1\over{\d-S_l}}{1\over{\d-S_{l-1}}}
  \ldots {1\over{\d-S_1}}\0\\\noal
&& +\sum_{l=1}^M \Bigl(a'_{N+l-1}+2a_{N+l-1}S_l\Bigl)'
{1\over{\d-S_l}}{1\over{\d-S_{l-1}}}\ldots {1\over{\d-S_1}}\label{c2}\\\noal
&& +\sum_{l=1}^M\sum_{k=1}^{l} a_{N+l-1}
{1\over{\d-S_l}}\ldots{1\over{\d-S_{k}}}
\bigg({2\over{N}}a_1'+2S_kS'_k-S_k^{''}\bigg)
 {1\over{\d-S_{k}}}\ldots {1\over{\d-S_1}}\0\\\noal
&& +\sum_{l=2}^M\sum_{k=1}^{l-1} a_{N+l-1}
{1\over{\d-S_l}}\ldots{1\over{\d-S_{k+1}}}
\Bigl(2S'_k\Bigl)
 {1\over{\d-S_{k}}}\ldots {1\over{\d-S_1}}.\0
\b
The last term here is not in the standard form: we have to move
$S_k'$ to the left
\a
&& +\sum_{l=2}^M\sum_{k=1}^{l-1} a_{N+l-1}
{1\over{\d-S_l}}\ldots{1\over{\d-S_{k+1}}}
\Bigl(2S'_k\Bigl)
 {1\over{\d-S_{k}}}\ldots {1\over{\d-S_1}}\0\\\noal
&&=\sum_{l=2}^M\sum_{k=1}^{l-1} \Bigl(2a_{N+l-1}S_k'\Bigl)
{1\over{\d-S_l}}{1\over{\d-S_{l-1}}}\ldots {1\over{\d-S_1}}\0\\\noal
&& +\sum_{l=2}^M\sum_{k=1}^{l-1} 2a_{N+l-1}[
{1\over{\d-S_l}}\ldots{1\over{\d-S_{k+1}}}, S_k']
 {1\over{\d-S_{k}}}\ldots {1\over{\d-S_1}}\label{c3}\\\noal
&&=\sum_{l=2}^M\Bigl(\sum_{k=1}^{l-1} 2a_{N+l-1}S_k'\Bigl)
{1\over{\d-S_l}}{1\over{\d-S_{l-1}}}\ldots {1\over{\d-S_1}}\0\\\noal
&& +\sum_{l=2}^M\sum_{k=1}^{l-1} 2a_{N+l-1}
{1\over{\d-S_l}}\ldots{1\over{\d-S_{k+1}}}\Bigl(-2\sum_{r=1}^{k-1}S_r^{''}\Bigl)
 {1\over{\d-S_{k}}}\ldots {1\over{\d-S_1}}.\0
\b
Combining eqs.(\ref{c1}), (\ref{c2}) and (\ref{c3}), and
comparing with eq(\ref{2ndf}), we obtain eqs.(\ref{gkdv}--\ref{ppkdv}),
which are the first non--trivial flow equations in the hierarchy (\ref{gkdvh}).

\section{Some simple $W(N, M)$ algebras}

\setcounter{equation}{0}
\setcounter{subsection}{0}
\setcounter{footnote}{0}

In this Appendix we give explicit expressions for a few simple
 $W(N, M)$ algebras.

\vskip3mm
\noindent
\underline{\bf $W(1,2)$ algebra}
\vskip2mm

\a
&&\{a_1, a_1\}_2=(a_1\d+\d a_1)\delta(x-y),
\quad\quad\{a_1, a_2\}_2=(a_2\d+2\d a_2)\delta(x-y),\0\\
&&\{a_1, S_1\}_2=(\d^2+S_1\d)\delta(x-y),
\quad\quad\{a_1, S_2\}_2=(2\d^2+S_2\d)\delta(x-y),\0\\
&&\{a_2, a_2\}_2=[(2a'_2+4a_2S_2-2a_2S_1)\d
+a^{''}_2+(2a_2S_2-a_2S_1)']
\delta(x-y),\label{w31}\\
&&\{a_2, S_2\}_2=\Bigl(a_1+(\d+S_2)(\d+S_2-S_1)\Bigl)
\delta'(x-y),\0\\
&&\{a_2, S_1\}_2=0,\quad
  \{S_i, S_j\}_2=(\delta_{ij}+1)\delta'(x-y),\quad i,j=1,2.\0
\b
The associated Lax operator is
\a
L=\d+a_1{1\over{\d-S_1}}+a_2{1\over{\d-S_2}}{1\over{\d-S_1}}.
\label{l02}
\b
This algebra generates the $W_\infty$ algebra through the
transformation (\ref{ulas}). For this reason it is called the 4--boson
representation of $W_\infty$ algebra.

\vskip3mm
\noindent
\underline{\bf $W(2,1)$ algebra}
\vskip2mm

\a
&&\{a_1, a_1\}=\Bigl(\ha\d^3+a_1\d+\d a_1\Bigl)\dxy,\0\\
&&\{a_2, a_2\}=\Bigl(\d^2a_2-a_2\d^2+
  2a_2S_2\d+2\d a_2S_2\Bigl)\dxy,\label{w32}\\
&&\{a_1, a_2\}=(a_2\d+2\d a_2)\dxy,\qquad
\{a_1, S_2\}=({3\over2}\d^2+S_2\d)\dxy,\0\\
&&\{a_2, S_2\}=\Bigl(a_1+(\d+S_2)^2\Bigl)\ddxy,\qquad
\{S_2, S_2\}={3\over2}\ddxy.\0
\b
The associated scalar Lax operator is
\a
L=\d^2+a_1+a_2{1\over{\d-S_2}}.\label{l11}
\b

\vskip3mm
\noindent
\underline{\bf $W(3, 0)=W_3$ algebra}
\vskip2mm

The $W(3,0)$ algebra is nothing but the $W_3$ algebra
\a
&&\{a_1, a_1\}=(2\d^3+a_1\d+\d a_1)\delta(x-y),\0\\
&&\{a_1, a_2\}=\Big(a_2\d+2\d a_2-\d^2 a_1-\d^4\Big)\delta(x-y),\label{w3}\\
&&\{a_2, a_2\}=[2a'_2\d+a^{''}_2-\frac{2}{3}(a_1+\d^2)(\d a_1+\d^3)]
\delta(x-y).\0
\b
The associated Lax operator is
\a
L=\d^3+a_1\d+a_2.\label{l20}
\b

\vskip3mm
\noindent
\underline{\bf $W(1,3)$ algebra}
\vskip2mm

\a
\{a_1, a_1\}&=&(a_1\d+\d a_1)\dxy,\qquad
\{a_1, a_2\}=(a_2\d+2\d a_2)\dxy,\0\\
\{a_1, a_3\}&=&(a_3\d+3\d a_3)\dxy,\qquad
  \{a_1, S_1\}=(\d^2+S_1\d)\dxy,\0\\
\{a_1, S_2\}&=&(2\d^2+S_2\d)\dxy,\qquad
  \{a_1, S_3\}=(3\d^2+S_3\d)\dxy,\0\\
\{a_2, a_2\}&=&\Bigl(\d^2a_2-a_2\d^2+2a_3\d+2\d a_3+
a_2(2S_2-S_1)\d+\d a_2(2S_2-S_1)\Bigl)\dxy,\0\\
\{a_2, a_3\}&=&\Bigl(3\d^2a_3-a_3\d^2-2S_1\d a_3-a_3\d S_1+2S_2\d a_3
   -\d a_3S_2\0\\
 &+&2a_3S_3\d+3\d a_3S_3\Bigl)\dxy,\label{w41}\\
\{a_2, S_1\}&=&0,\qquad
\{a_2, S_2\}=\Bigl(a_1+(\d+S_2)(\d+S_2-S_1)\Bigl)\ddxy,\0\\
\{a_2, S_3\}&=&\Bigl(a_1+3\d^2+(3S_3+S_2-2S_1)\d+
   (S_3^2-S_1S_3-S_1'-S_2'+2S_3')\Bigl)\ddxy,\0\\
\{a_3, a_3\}&=&\Bigl(a_3\d^3+\d^3 a_3+a_3\d a_1+a_1\d a_3
   +a_3(S_1+S_2-3S_3)\d^2\0\\
 &-&\d^2 a_3(S_1+S_2-3S_3)+
 a_3(3S_3^2+S_1S_2-2S_1S_3-2S_2S_3+S_1'-3S_3')\d\0\\
 &+&\d a_3(3S_3^2+S_1S_2-2S_1S_3-2S_2S_3+S_1'-3S_3')\Bigl)\ddxy\0\\
\{a_3, S_3\}&=&\Bigl(a_1\d+a_2+a_1(S_3-S_2)+(\d+S_3)(\d+S_3-S_1)(\d+S_3-S_2)
   \Bigl)\ddxy,\0\\
\{a_3, S_i\}&=&0,\quad i=1,2;\qquad
   \{S_i, S_j\}=(\delta_{ij}+1)\ddxy, \quad i,j=1,2,3.\0
\b
The Lax operator is
\a
L=\d+a_1{1\over{\d-S_1}}+a_2{1\over{\d-S_2}}{1\over{\d-S_1}}
  +a_3{1\over{\d-S_3}}{1\over{\d-S_2}}{1\over{\d-S_1}}.\label{l03}
\b
This algebra also generates the $W_\infty$ algebra, and is called the
$6$--field representation of the $W_\infty$ algebra.

\vskip3mm
\noindent
\underline{\bf $W(2,2)$ algebra}
\vskip2mm

\a
&&\{a_1, a_1\}=\Bigl(\ha\d^3+a_1\d+\d a_1\Bigl)\dxy,\0\\
&&\{a_1, a_2\}=(a_2\d+2\d a_2)\dxy,\qquad
  \{a_1, a_3\}=(a_3\d+3\d a_3)\dxy,\0\\
&&\{a_1, S_2\}=({3\over2}\d^2+S_2\d)\dxy,\qquad
  \{a_1, S_3\}=({5\over2}\d^2+S_3\d)\dxy,\0\\
&&\{a_2, a_2\}=\Bigl(\d^2a_2-a_2\d^2+2a_3\d+2\d a_3+
2a_2S_2\d+2\d a_2S_2\Bigl)\dxy,\0\\
&&\{a_2, S_2\}=\Bigl(a_1+(\d+S_2)^2\Bigl)\ddxy,\label{42}\\
&&\{a_2, S_3\}=\Bigl(a_1+3\d^2+(S_2+3S_3)\d+
   (S_3^2+2S_3'-S_2')\Bigl)\ddxy,\0\\
&&\{a_2, a_3\}=\Bigl(3\d^2a_3-a_3\d^2+2S_2\d a_3
   -\d a_3S_2+2a_3S_3\d+3\d a_3S_3\Bigl)\dxy,\0\\
&&\{a_3, a_3\}=\Bigl(a_3\d^3+\d^3 a_3+a_3\d a_1+a_1\d a_3+a_3(S_2-3S_3)\d^2
  -\d^2 a_3(S_2-3S_3)\0\\
&&\qquad+ a_3(3S_3^2-2S_2S_3-3S_3')\d+\d a_3(3S_3^2-2S_2S_3-3S_3')\Bigl)
  \ddxy\0\\
&&\{a_3, S_3\}=\Bigl(a_1\d+a_2+a_1(S_3-S_2)+(\d+S_3)^2(\d+S_3-S_2)
   \Bigl)\ddxy,\0\\
&&\{a_3, S_2\}=0,\qquad
   \{S_i, S_j\}=(\delta_{ij}+\ha)\ddxy, \quad i,j=2,3.\0
\b
In this case the scalar Lax operator is
\a
L=\d^2+a_1+a_2{1\over{\d-S_2}}+a_3{1\over{\d-S_3}}{1\over{\d-S_2}}.\label{l12}
\b

\vskip3mm

      \noindent
\underline{\bf $W(3,1)$ algebra}
\vskip2mm

\a
&&\{a_1, a_1\}=(2\d^3a_1\d+\d a_1)\dxy,\0\\
&&\{a_1, a_2\}=(a_2\d+2\d a_2-\d^2a_1-\d^4)\dxy,\0\\
&&\{a_1, a_3\}=(a_3\d+3\d a_3)\dxy,\qquad
  \{a_1, S_3\}=(2\d^2+S_3\d)\dxy,\0\\
&&\{a_2, a_2\}=\Bigl(\d^2a_2-a_2\d^2+2a_3\d+2\d a_3-{2\over3}
  (a_1+\d^2)\d(\d^2+a_1)\Bigl)\dxy,\0\\
&&\{a_2, a_3\}=\Bigl(3\d^2a_3-a_3\d^2
   +2a_3S_3\d+3\d a_3S_3\Bigl)\dxy,\label{w43}\\
&&\{a_2, S_3\}=\Bigl({2\over3}a_1+{8\over3}\d^2+3S_3\d+
   (S_3^2+2S_3')\Bigl)\ddxy,\0\\
&&\{a_3, a_3\}=\Bigl(a_3\d^3+\d^3 a_3+a_3\d a_1+a_1\d a_3-3a_3S_3\d^2
   +3\d^2 a_3S_3\0\\
&&\qquad+ 3a_3(S_3^2-S_3')\d
+3\d a_3(S_3^2-S_3')\Bigl)\ddxy\0\\
&&\{a_3, S_3\}=\Bigl(a_2+a_1(\d+S_3)+(\d+S_3)^3\Bigl)\ddxy,\0\\
&&\{S_3, S_3\}={4\over3}\ddxy.\0
\b
with Lax operator
\a
L=\d^3+a_1\d+a_2+a_3{1\over{\d-S_3}}.\label{l21}
\b

\vskip3mm
\noindent
\underline{\bf $W(4,0)=W_4$ algebra}
\vskip2mm

At last let us give here the $W_4$ algebra
\a
&&\{a_1, a_1\}=(5\d^3a_1\d+\d a_1)\dxy,\0\\
&&\{a_1, a_2\}=(a_2\d+2\d a_2-2\d^2a_1-5\d^4)\dxy,\0\\
&&\{a_1, a_3\}=\Bigl(a_3\d+3\d a_3+{3\over2}(\d^5+\d^3 a_1-\d^2 a_2)\Bigl)
     \dxy,\0\\
&&\{a_2, a_2\}=\Bigl(\d^2a_2-a_2\d^2+2a_3\d+2\d a_3-a_1\d a_1-2a_1\d^3
  -2\d^3 a_1-6\d^5\Bigl)\dxy,\label{w4}\\
&&\{a_2, a_3\}=\Bigl(3\d^2a_3-a_3\d^2+\ha(a_1+4\d^2)\d(\d a_1-a_2+\d^3)
    \Bigl)\dxy,\0\\
&&\{a_3, a_3\}=\Bigl(a_3\d^3+\d^3 a_3+a_3\d a_1+a_1\d a_3\0\\
&&\qquad+ {3\over4}(a_2+a_1\d+\d^3)\d(\d^3+\d a_1-a_2)\Bigl)\ddxy.\0
\b
It is well--known that this algebra is associated with the scalar Lax
operator
\a
L=\d^4+a_1\d^2+a_2\d+a_3.\label{l30}
\b

\section{Some $W(N, M)$ algebras in the $(q, r)$--gauge}

\setcounter{equation}{0}
\setcounter{subsection}{0}
\setcounter{footnote}{0}

\vskip3mm
\noindent
\underline{\bf $W(2,1)$ algebra}
\vskip2mm

\a
&&\{a_1, a_1\}=\Bigl(\ha\d^3+a_1\d+\d a_1\Bigl)\dxy,\0\\
&&\{a_1, q\}=(q\d+\ha\d q)\dxy,\qquad
\{a_1, r\}=(r\d+\ha\d r)\dxy,\0\\
&&\{q, q\}=-\frac{3}{2}q\d\inv q\dxy,\qquad
\{r, r\}=-\frac{3}{2}r\d^{-1} r\dxy,\0\\
&&\{q, r\}=\Bigl(\d^2+a_1+\frac{3}{2}q\d^{-1}r\Bigl)\dxy.\0
\b
The associated scalar Lax operator
\a
L=\d^2+a_1+q\d^{-1}r.\0
\b

\vskip3mm
\noindent
\underline{\bf $W(2,2)$ algebra}
\vskip2mm

\a
&&\{a_1, a_1\}=\Bigl(\ha\d^3+a_1\d+\d a_1\Bigl)\dxy,\0\\
&&\{a_1, r_1\}=(r_1\d+\ha\d r_1)\dxy,\qquad
\{a_1, r_2\}=(r_2\d+\frac{3}{2}\d r_2)\dxy,\0\\
&&\{a_1, q_1\}=(q_1\d+\ha\d q_1)\dxy,\qquad
\{a_1, q_2\}=(q_2\d+\frac{1}{2}\d q_2)\dxy,\0\\
&&\{r_i, r_j\}=-(\delta_{ij}+\ha)r_r\d^{-1}r_j\dxy,\qquad i,j=1,2,\qquad
\{q_2, r_1\}=\frac{1}{2}q_2\d^{-1}r_1\dxy,\0\\
&&\{q_1, r_1\}=\Bigl(\d^2+a_1+\frac{3}{2}q_1\d^{-1}r_1\Bigl)\dxy,\0\\
&&\{q_2, r_2\}=\Bigl(q_1r_1+\frac{a_1}{r_1}\d r_1+\d^2\frac{1}{r_1}\d r_1
+\frac{3}{2}q_2\d^{-1}r_2\Bigl)\dxy,\0\\
&&\{q_1, r_2\}=\Bigl(\frac{r_2}{r_1}a_1+\frac{r_2}{r_1}\d^2
+\frac{1}{r_1}\d r_2\d+\frac{2}{r_1}\d^2 r_2-\frac{1}{r_1^2}\d r_1\d r_2
+\frac{r_2}{r_1}\d\frac{1}{r_1}\d r_1-\frac{r_2}{r_1^2}
+\ha q_1\d^{-1}r_2\Bigl)\dxy,\0\\
&&\{q_1, q_1\}=\Bigl(-\frac{3}{2}q_1\d^{-1}q_1+\frac{2q_2}{r_1}\d
\frac{r_2}{r_1}+\frac{2r_2}{r_1}\d\frac{q_2}{r_1}\Bigl)\dxy\0\\
&&\{q_1, q_2\}=\Bigl(\frac{1}{r_1}\d^2 q_2-\frac{2}{r_1}\d q_2\d
-\ha q_1\d^{-1}q_2-\frac{q_2}{r_1}a_1\Bigl)\dxy,\0\\
&&\{q_2, q_2\}=-\frac{3}{2}q_2\d^{-1}q_2\Bigl)\dxy.\0
\b
In this case the scalar Lax operator is
\a
L=\d^2+a_1+q_1\d^{-1}r_1+q_2\d^{-1}\frac{r_2}{r_1}\d^{-1}r_1.\0
\b

\section{Identities satisfied by the polynomials $e_k$}

\setcounter{equation}{0}
\setcounter{subsection}{0}
\setcounter{footnote}{0}

We devote this Appendix to proving the identities (\ref{pks1}) and
(\ref{dsp}--\ref{dspdsp}).

\vskip0.2cm
\noindent
\underline{Proof of eq.(\ref{pks1})}
\vskip0.2cm

Let us define
\a
G(p; S_1,S_2.\ldots,S_M)=\frac{1}{(p-S_1)(p-S_2)\ldots(p-S_M)},\0
\b
which is the generating function of the symmetric polynomials $e_k$,
\a
G=\sum_{l=0}^\infty e_l p^{-l-M}.\label{gpl}
\b
Taking one time derivative with respect to, say, $S_1$ for $G$,
then expanding it in the powers of $p$, we obtain (\ref{pks1}).

Likewise we can prove eqs.(\ref{dsp}--\ref{p'}).

\vskip0.2cm
\noindent
\underline{Proof of eq.(\ref{pij-+})}
\vskip0.2cm

\def\talpha{\tilde \alpha}
\def\tbeta{\tilde \beta}
\a
{\rm LHS}&\stackrel{\talpha=\alpha-\beta}{\longrightarrow}&
\sum_{l=1}^M\sum_{\beta=0}^{i-1}\sum_{\talpha=1}^{i-\beta}
 S^\beta_l e_{i-\beta-\talpha}e_{j+\talpha}\0\\
&=&\sum_{\talpha=1}^i\sum_{\beta=0}^{i-\talpha}
 S^\beta_l e_{i-\beta-\talpha}e_{j+\talpha}\0\\
&=&\sum_{\alpha=1}^i(i-\alpha+M)
 e_{i-\alpha}e_{j+\alpha}={\rm RHS}.\0
\b
In the last step we have used eq.(\ref{sumsp}).
The proof of eq.(\ref{pi'j-+}) is the same as this.

\vskip0.2cm
\noindent
\underline{Proof of eq.(\ref{dspdsp})}
\vskip0.2cm

Since eq.(\ref{dspdsp}) is an operatorial equation, we can cut it into
two pieces.
\ai
&&\sum_{l=1}^M\Bigl(\frac{\d}{\d S_l}e_{i+1}\Bigl)\Bigl(
 \frac{\d}{\d S_l}e_{j+1}\Bigl)
=(j+M)e_{i}e_{j}
+\sum_{l=1}^i
(j-i+2l)e_{i-l}e_{j+l}.\label{dspdsp1}\\
&&\sum_{l=1}^M\Bigl(\frac{\d}{\d S_l}e_{i+1}\Bigl)\Bigl(
 \frac{\d}{\d S_l}e_{j+1}\Bigl)'
=(j+M)e_{i}e'_{j}
+\sum_{l=1}^i\Bigl(le_{j+l}e'_{i-l}
+(j-i+l)e_{i-l} e'_{j+l}\Bigl).\label{dspdsp2}
\bj
Now let us consider first eq.(\ref{dspdsp1})
\a
{\rm LHS}&=&\sum_{l=1}^M\sum_{\alpha=0}^i\sum_{\beta=0}^je_{i-\alpha}
e_{j-\beta}S_l^{\alpha+\beta}
=\sum_{l=1}^M\sum_{\alpha=0}^i\sum_{\beta=\alpha}^{j+\alpha}e_{i-\alpha}
e_{j+\alpha-\beta}S_l^{\beta}\0\\
&=&\sum_{l=1}^M\sum_{\alpha=0}^i\Bigl(\sum_{\beta=0}^{j+\alpha}
-\sum_{\beta=0}^{\alpha-1}\Bigl)e_{i-\alpha}
e_{j+\alpha-\beta}S_l^{\beta}\0\\
&=&\sum_{\alpha=0}^i(j+\alpha+M)e_{i-\alpha}e_{j+\alpha}
-\sum_{l=1}^M\sum_{\beta=0}^{i-1}
\sum_{\alpha=\beta+1}^i e_{i-\alpha}
e_{j+\alpha-\beta}S_l^{\beta}\0\\
&=&\sum_{\alpha=0}^i(j+\alpha+M)e_{i-\alpha}e_{j+\alpha}
-\sum_{\alpha=1}^i(i-\alpha+M)e_{i-\alpha}e_{j+\alpha}={\rm RHS}.\0
\b
The LHS of (\ref{dspdsp2}) can be divided into two pieces the first being
\a
A&=&\sum_{k,l=1}^M\sum_{\al=0}^i\sum_{\beta=0}^{j-1}
\sum_{\mu=0}^{j-\beta-1}S_k'S_k^\mu S_l^{\al+\beta}
e_{i-\al}e_{j-\beta-\mu-1}\0\\
&=&\sum_{k=1}^M\sum_{\mu=0}^{j-1}S_k'S_k^\mu
\sum_{l=1}^M\sum_{\al=0}^i\sum_{\beta=\al}^{j+\al-1}
S_l^{\al+\beta}
e_{i-\al}e_{j-\beta-\mu-1},\0
\b
Splitting one of the above summations as follows
\a
\sum_{\beta=\al}^{j+\al-1}=\sum_{\beta=0}^{j+\al-1}-\sum_{\beta=0}^{\al-1},\0
\b
and using eqs.(\ref{p'}, \ref{pij-+}), we get
\a
A&=&\sum_{l=1}^M\sum_{\mu=0}^{j-1}\Bigl(\sum_{\al=0}^i(j+\al-\mu-1+M)
-\sum_{\al=1}^i(i-\al+M)\Bigl)e_{i-\al}e_{j+\al-\mu-1}
S_k'S_k^\mu \0\\
&=&(j+M)e_iP'_j+\sum_{\al=1}^i(j-i+\al)e_{i-\al}e_{j+\al}
 -\sum_{k=1}^M\sum_{\mu=0}^{j-1}(\mu+1)e_i e_{j-\mu-1}S_k^\mu S_k'\0\\
&+&\sum_{k=1}^M\sum_{\al=1}^i\Bigl(\sum_{\beta=0}^{j-1}(\al-\beta-1)
 -\sum_{\beta=j}^{j+\al-1}(\al-\beta-1)\Bigl)
 e_{i-\al}e_{j+\al-\beta-1}S_k^\mu S_k'.\label{111}
\b
The other term in (\ref{dspdsp2}) is
\a
B&=&\sum_{k=1}^M\sum_{\al=0}^i\sum_{\beta=0}^j
\beta e_{i-\al}e_{j-\beta}S_k^{\al+\beta-1} S_k'\0\\
&=&\sum_{k=1}^M\sum_{\beta=0}^{j-1}(\beta+1)e_i e_{j-\beta-1}S_k^\beta S_k'
 +\sum_{k=1}^M\sum_{\al=1}^i\sum_{\beta=\al}^{j+\al-1}(\beta-\al+1)
 e_{i-\al}e_{j+\al-\beta-1}S_k^\beta S_k'.\label{112}
\b
Combining (\ref{111}) and (\ref{112}) and making use of
eq.(\ref{pi'j-+}), we obtain
\a
A+B&=&(j+M)e_ie'_j+\sum_{\al=1}^i(j-i+\al)e_{i-\al}e'_{j+\al}\0\\
&+&\sum_{k=1}^M\sum_{\al=1}^i\sum_{\beta=0}^{\al-1}(\al-\beta-1)
 e_{i-\al}e_{j+\al-\beta-1}S_k^\beta S_k'\0\\
&-&\sum_{k=1}^M\sum_{\al=1}^i\sum_{\beta=j}^{j+\al-1}(j-i+2\al-\beta-1)
 e_{i-\al}e_{j+\al-\beta-1}S_k^\beta S_k'\0\\
&=&(j+M)e_i e'_j+\sum_{\al=1}^i\Bigl(\al e'_{i-\al}e_{j+\al}
 +(j-i+\al)e_{i-\al}e'_{j+\al}\Bigl),\0
\b
The last term in the intermediate expression vanishes as one can see in the
following
way. Let us split the term into two parts
\a
\sum_{k=1}^M\sum_{\al=1}^i\sum_{\beta=j}^{j+\al-1}
(j+\al-\beta-1)e_{i-\al}e_{j+\al-\beta-1}S_k^\beta S_k',\qquad
\sum_{k=1}^M\sum_{\al=1}^i\sum_{\beta=j}^{j+\al-1}
(\al-i)e_{i-\al}e_{j+\al-\beta-1}
S_k^\beta S_k',\0
\b
and redefine the summation parameter for the first part as follows
\a
\talpha=i-j-\al+\beta+1,\qquad \tbeta=\beta,\0
\b
After changing the order of the summations the first part it is exactly
the same as the second except for the different sign. This completes
our proof of eq.(\ref{dspdsp2}).

\end{document}